\documentclass[prd,preprint,tightenlines,showpacs,preprintnumbers,nofootinbib,eqsecnum,superscriptaddress]{revtex4}

\usepackage[T1]{fontenc} 
\usepackage{amsfonts,amssymb,amscd,amsmath}
\usepackage[dvipsnames]{xcolor}
\usepackage{graphicx}
\usepackage{verbatim}
\newcommand{\rr}{\mbox{\boldmath $r$}}
\newcommand{\bPhi}{\mbox{\boldmath $\Phi$}}

\newcommand{\be}{\mbox{\boldmath $e$}}

\newcommand{\bsigma}{\mbox{\boldmath $\sigma$}}

\newcommand{\qv}{\mbox{\boldmath $q$}}

\newcommand{\kv}{\mbox{\boldmath $k$}}

\newcommand{\dd}{\, \mathrm{d}}

\graphicspath{{figures/}}
\begin{document}
\title{\boldmath Dilepton angular distributions in the color-dipole $S$-matrix framework}
\author{Yan B. Bandeira}
\email{yan.bandeira@ufpel.edu.br}
\affiliation{Institute of Physics and Mathematics, Federal University of Pelotas,\\Postal Code 354, 96010-900, Pelotas, RS, Brazil}
\affiliation{Institute of Nuclear Physics Polish Academy of Sciences,\\PL-31-342, Kraków, Poland}

\author{Victor P. Gonçalves}
\email{barros@ufpel.edu.br}
\affiliation{Institute of Physics and Mathematics, Federal University of Pelotas,\\Postal Code 354, 96010-900, Pelotas, RS, Brazil}

\author{Wolfgang Sch\"afer}
\email{Wolfgang.schafer@ifj.edu.br}
\affiliation{Institute of Nuclear Physics Polish Academy of Sciences,\\PL-31-342, Kraków, Poland}

\begin{abstract}
   The dilepton production at forward rapidities in hadronic collisions at the LHC energies is considered and the helicity density matrix elements of gauge bosons that enter the Drell--Yan angular coefficients associated with the gauge boson decay ($G = \gamma^*, Z^0, W$) are derived using the color-dipole $S$ - matrix framework.
   We show results for $pp$ collisions at $\sqrt{s} = 14$ TeV for the nonvanishing helicity density matrix elements in the rapidity range of  $2.0 \le y \le 4.0$ of the gauge boson. We consider different models for the proton unintegrated gluon distributions, among others a solution of the Balitsky--Kovchegov equation. We compare the associated predictions with those obtained disregarding the non-linear term in the evolution equation. In addition, the Lam--Tung relation is discussed.
\end{abstract}
\maketitle
\flushbottom

\section{Introduction}

The dilepton production in the hadronic interactions -- the Drell--Yan  process -- has since long been one of the main tools for studying the partonic structure of hadrons and properties of Standard--Model gauge bosons \cite{ATLAS:2012au,ATLAS:2016rnf,CMS:2015cyj,LHCb:2022tbc}.
While for dilepton production via virtual photons, there are four parity conserving structure functions \cite{Lam:1978pu}, for the case of electroweak gauge bosons $W^\pm, Z^0$, the lepton angular distributions give access to in total nine independent structure functions \cite{Korner:1990im,Mirkes:1992hu,Lyubovitskij:2024civ}.

Here we expand on recent work  \cite{Schafer:2016qmk,Bandeira:2024zjl,Bandeira:2024jjl} relevant for the production of gauge bosons $G =\gamma^*, Z^0, W^\pm$ at forward rapidities, deriving the full polarization density matrix of gauge bosons produced by the $q \to q' G, \bar q \to \bar q' G$ transitions in the color field of the target. 
The study of dilepton production at forward rapidities is considered a promising probe of quantum chromodynamics (QCD) at high energies by giving access to the gluon distribution or its unintegrated transverse--momentum dependent generalizations at small fractions of their momentum fraction $x$. 

Over the last decades, several groups have estimated the transverse and rapidity differential distributions considering proton-proton ($pp$) and proton-nucleus ($pA$) collisions at the RHIC and LHC energies assuming different approaches for the description of the cross-section and distinct models for the treatment of the possible non-linear effects, which are expected to modify the QCD dynamics at small-$x$. Many of these calculations make use of the color--dipole approach \cite{Nikolaev:1990ja} extended to the Drell--Yan process in \cite{Kopeliovich:1995an,Brodsky:1996nj}, see for example \cite{Kopeliovich:2000fb,Raufeisen:2002zp,Kopeliovich:2001hf,Betemps:2003je,Golec-Biernat:2010dup,Schafer:2016qmk,Basso:2015pba,Basso:2016ulb}. 

\begin{figure}[t]
    \centering
    \includegraphics[width=0.8\textwidth]{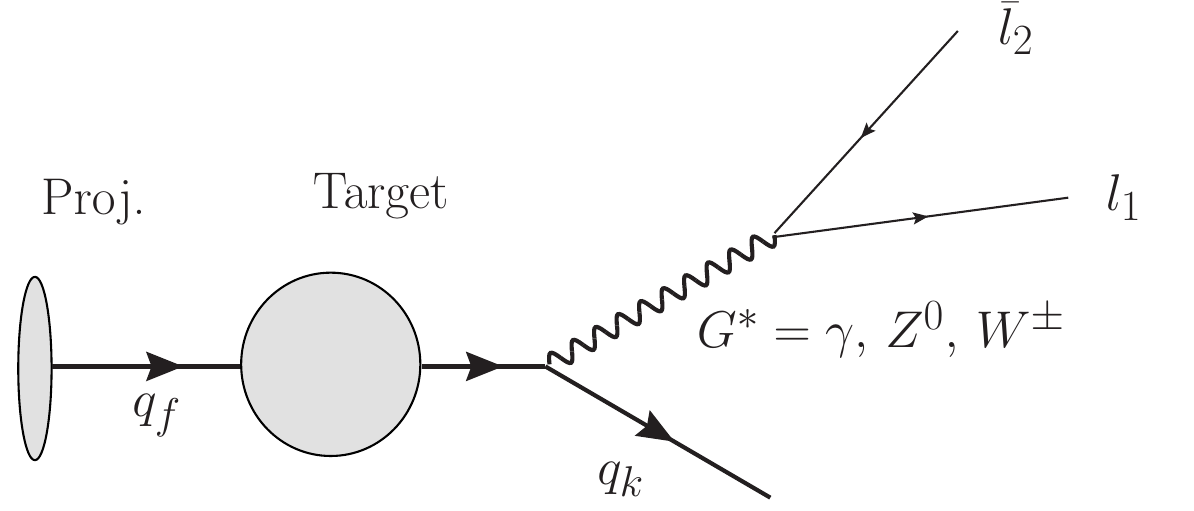}
    \caption{One of two diagrams for the Drell--Yan process in the hybrid factorization formalism. In the high energy limit, the gauge boson vertex can occur before or after interaction with the target fields.}
    \label{fig:diagram}
\end{figure}

One of the largely used approaches, which naturally  is also part of the color-dipole approach, is the hybrid factorization formalism proposed for the Drell--Yan process in \cite{Gelis:2002fw}. 
In this formalism, for the dilute projectile the DY process is also considered as an electroweak gauge boson bremsstrahlung off a fast projectile quark propagating
through the low-$x$ color field of the target,  as illustrated in Fig. \ref{fig:diagram}, but 
the quark-target cross-section,  which can include the non-linear QCD effects is treated for example in the color--glass condensate approach \cite{Ducloue:2017kkq,Marquet:2019ltn,Stasto:2012ru,Kang:2012vm}. 
Other approaches based on tranverse momentum--dependent parton distributions use on--shell matrix elements, see \cite{Piloneta:2024aac} or off-shell generalizations \cite{Nefedov:2012cq,Motyka:2016lta,Nefedov:2018vyt,Celiberto:2018muu}.
Finally, high perturbative orders have been achieved in collinear factorization \cite{Gauld:2017tww}.

Recently, in Refs.  \cite{Bandeira:2024zjl,Bandeira:2024jjl}, we have derived the general formulae for the inclusive electroweak gauge boson production at forward rapidities using the $S$-matrix framework proposed in Refs. \cite{Nikolaev:1994de,Nikolaev:1995ty,Nikolaev:2003zf,Nikolaev:2004cu,Nikolaev:2005dd,Nikolaev:2005zj,Nikolaev:2005ay,Nikolaev:2005qs} and demonstrated that it reduces to those used in Refs. \cite{Kopeliovich:2007yva,Kopeliovich:2009yw,SampaiodosSantos:2020lte,Gelis:2002ki,Jalilian-Marian:2012wwi,Ducloue:2017kkq,Goncalves:2020tvh,Lima:2023dqw,Kopeliovich:2000fb,Raufeisen:2002zp,Kopeliovich:2001hf,Betemps:2004xr,Betemps:2003je,Golec-Biernat:2010dup,Ducati:2013cga,Schafer:2016qmk,Ducloue:2017zfd,Gelis:2002fw,Baier:2004tj,Stasto:2012ru,Kang:2012vm,Basso:2016ulb,Basso:2015pba,Marquet:2019ltn} to estimate the real photon and $Z^0$ production in the appropriate limits and representations.  In addition,  the cross section for the $W^{\pm}$ production in the hybrid factorization formalism was derived for the first time. 

A natural next step is to expand the results derived in Refs. \cite{Bandeira:2024zjl,Bandeira:2024jjl} for less inclusive observables, which provide additional information about the hadronic structure, as e.g., the angular distribution of the leptons pairs ($\ell_1 \bar{\ell}_2$).
The important roles of the lepton angular distribution in understanding the mechanisms for $W$, $Z$ and $\gamma$ boson production in hadronic collisions were pointed out e.g. in Refs. \cite{Mirkes:1992hu,Mirkes:1994eb,Boer:2006eq}. Differently from  $\gamma$ and $Z$ boson production, where both $\ell_1$ and $\bar{\ell}_2 = \bar{\ell}_1$ decay
products are detected, only the charged lepton from $W$
boson decay is measured. It implies that distinct experimental uncertainties are encountered in the measurements
of lepton angular distributions associated with the production of the different gauge bosons. Another important difference is that these distributions involve different couplings. In particular,  $W$ and $Z$
boson productions involve different parity-violating couplings, which makes instructive to compare the corresponding lepton angular distributions. 

In this paper, we will focus on the calculation of the angular distributions of leptonic pairs ($\ell_1 \bar{\ell}_2$) at forward rapidities and LHC energies, expanding our recent work \cite{Bandeira:2024zjl,Bandeira:2024jjl} and generalizing the results derived in Ref.  \cite{Schafer:2016qmk} for the cases where the dilepton system is produced by a $Z^0$ or a $W^{\pm}$ gauge boson.
Our goal is to derive  all  density matrix elements need to estimate the angular coefficients using the color-dipole $S$-matrix framework. As we will demonstrate, the resulting expressions will be linearly dependent of the target unintegrated gluon distribution (UGD), which is sensitive to the description of the QCD dynamics, and are valid for $pp$ and $pA$ collisions. In our analysis, we will  calculate the angular coefficients in $pp$ collisions at $\sqrt{s} = 14$ TeV, assuming that the dileptons are produced in the rapidity range covered by the LHCb detector, and considering distinct models for the proton UGD. In particular, we will consider two models that take into account of non-linear effects in the QCD dynamics and will compare with those derived disregarding these effects.

This paper is organized as follows. In the next section, we derive the expressions for the gauge boson polarization density matrix elements in terms of the structure functions present in the decomposition of the hadronic tensor. In particular, we will demonstrate that distinctly from the case of  virtual photons,  where there are four parity conserving structure functions \cite{Lam:1978pu}, for the case of electroweak gauge bosons $W^\pm, Z^0$, the lepton angular distributions give access to nine independent structure functions \cite{Korner:1990im,Mirkes:1992hu,Lyubovitskij:2024civ}. In section \ref{sec:DMEs} we will present the derivation of the density matrix elements using the color - dipole $S$ - matrix framework for the generic $q \rightarrow qG$ process, with $G = \gamma$, $Z$ or a $W$ boson. We will demonstrate that the results derived in Ref.  \cite{Schafer:2016qmk} are recovered for $G = \gamma$.   Our predictions for the distinct density matrix elements will be presented in section \ref{sec:res}, considering the dilepton production at forward rapidities ($2.0 \le y \le 4.0$) in $pp$ collisions at $\sqrt{s} = 14$ TeV. In order to estimate the impact of the non - linear effects on the density matrix elements, we will present the results associated with distinct models for the proton UGD. Finally, in section \ref{sec:sum}, we will summarize our main results and conclusions. For completeness of our study, three appendices are included, where we present the explicit expressions for the LCWF, the relation of the angular coefficients present in our calculation with other conventions found in the literature, as well as the relation between the structure functions derived in distinct coordinate frames.

\section{Structure functions and the gauge boson polarization density matrix}

The cross - section for the Drell - Yan process in $pp$ collisions at the center - of - mass energy $\sqrt{s}$ can be expressed in terms of the leptonic $L^{\mu\nu}$ and hadronic tensors $W_{\mu\nu}$ as follows
\begin{equation}
    (2\pi)^4 \, \frac{d \sigma (pp \to \ell_1(l_1) \bar{\ell}_2 (l_2) + X)}{d^4 q} = \frac{1}{2s }\,\frac{W_{\mu\nu}L^{\mu\nu}}{(M^2 - M_G^2)^2 + \Gamma^2_G M_G^2} \,  d\Phi(q,l_1,l_2),
\end{equation}
where $q = l_1 + l_2$ is the four-momentum of the virtual gauge boson, $M^2 = q^2$ is the invariant mass of the lepton pair, $M_G$ is the gauge boson mass, $\Gamma_G$ is its total decay width and $d\Phi$ is the dilepton phase space.  We have adopted the relativistic Breit--Wigner form as appropriate for a narrow resonance. As emphasized e.g. in Ref. \cite{Mirkes:1992hu}, the lepton tensor $L_{\mu \nu}$ acts as an analyzer of the gauge boson polarization. In contrast,  $W_{\mu\nu}$ contains all the dynamical information on the gauge boson production. 


As there is no standardized notation for the relevant structure functions, we adopt the tensor decomposition and notational conventions recently
proposed in Ref.\cite{Lyubovitskij:2024civ} and write:
\begin{equation}\label{eq:hadronic=000020tensor}
    \begin{split}
W^{\mu\nu}=\  & \left(X^{\mu}X^{\nu}+Y^{\mu}Y^{\nu}\right)W_{T}+i\left(X^{\mu}Y^{\nu}
 -Y^{\mu}X^{\nu}\right)W_{T_{p}}    
 +Z^{\mu}Z^{\nu}W_{L}\\
& +\left(Y^{\mu}Y^{\nu}-X^{\mu}X^{\nu}\right)W_{\Delta\Delta}
  -\left(X^{\mu}Y^{\nu}+Y^{\mu}X^{\nu}\right)W_{\Delta\Delta_{p}} \\
&-\left(X^{\mu}Z^{\nu}+Z^{\mu}X^{\nu}\right)W_{\Delta}
  -\left(Y^{\mu}Z^{\nu}+Z^{\mu}Y^{\nu}\right)W_{\Delta_{p}} \\
&+i\left(Z^{\mu}X^{\nu}-X^{\mu}Z^{\nu}\right)W_{\nabla}
 +i\left(Y^{\mu}Z^{\nu}-Z^{\mu}Y^{\nu}\right)W_{\nabla_{p}}.
    \end{split}
\end{equation}
Here $X^\mu, Y^\mu, Z^\mu$ is a orthogonal basis of spacelike vectors, which fulfill $X^2 = Y^2 = Z^2 = -1$.
Together with the timelike unit vector $T^\mu = q^\mu/M, T^2 = 1$, they form a right handed, i.e. $Y^\mu = \epsilon^{\mu \nu \alpha \beta} T_\nu Z_\alpha X_\beta$,  basis of Minkowski space. 
In a given rest frame of the gauge boson they can be thought of having the canonical form
$T^\mu=(1,0,0,0), X^\mu=(0,1,0,0), Y^\mu = (0,0,1,0), Z^\mu = (0,0,0,1)$.
For useful covariant expressions of $X^\mu,Y^\mu,Z^\mu$ in terms of the four momenta of incoming hadrons and the gauge boson, which fix the orientation of axes with relation to these particle momenta, see e.g. \cite{Boer:2006eq}. 
In our discussion, we will adopt a convention, where $X^\mu, Z^\mu$ span the gauge--boson production plane, and $Y^\mu$ is the out-of plane direction.
The Gottfried--Jackson and Collins--Soper frames discussed below differ by a rotation in the production plane. Other frame choices are discussed e.g. in \cite{Faccioli:2022peq}.

The lepton tensor for massless leptons coupled to gauge bosons via $g_V \gamma_\mu - g_A \gamma_\mu \gamma_5$, with the respective vector and axial couplings $g_V, g_A$ reads
\begin{eqnarray}
    L_{\mu \nu} = 4(g_V^2 + g_A^2) \Big( l_{1\mu} l_{2\nu} + l_{2\mu} l_{1\nu} - (l_1 \cdot l_2) g_{\mu \nu} \Big) - 8 g_V g_A i \epsilon_{\mu \nu \alpha \beta} l_1^\alpha l_2^\beta  \,\,.
\end{eqnarray}
In a given rest frame of the lepton pair, we parameterize the four vectors of leptons, which we take to be massless, as
\begin{eqnarray}
    l_{1\mu} = \frac{M}{2} ( T_\mu + n_\mu) , \quad l_{2\mu} = \frac{M}{2} (T_\mu - n_\mu) \, , 
\end{eqnarray}
with
\begin{eqnarray}
    n_\mu = \sin \theta \cos \phi \,  X_\mu + \sin \theta \sin \phi \, Y_\mu + \cos \theta \, Z_\mu .
\end{eqnarray}
The lepton tensor then obtains the form
\begin{eqnarray}
    L_{\mu \nu} =
    2 M^2 (g_V^2 + g_A^2) \, {\cal L}_{\mu \nu}, 
\end{eqnarray}
with
\begin{eqnarray}
    {\cal L}_{\mu \nu} = - g_{\mu \nu} + T_\mu T_\nu - n_\mu n_\nu + i c_G  \, \epsilon_{\mu \nu \alpha \beta} T^\alpha n^\beta \, ,  
\end{eqnarray}
where
\begin{eqnarray}
    c_G = \frac{2 g_V g_A}{g_V^2 + g_A^2}.
\end{eqnarray}
Here, in the Standard Model $c_\gamma =0, c_W = 1, c_Z \approx 0.08$.
The lepton angular distributions defined via
\begin{eqnarray}
    \frac{d \sigma}{d^4q d\Omega} = \frac{d \sigma}{d^4 q} \, \frac {dN}{d\Omega}
\end{eqnarray}
can then be obtained after contraction with the hadronic tensor in a straightforward manner as:
\begin{eqnarray}
    \frac{dN}{d \Omega} = \frac{3}{8 \pi} \, \frac{W^{\mu \nu} {\cal L}_{\mu \nu}}{2W_T + W_L} \,\,.
\end{eqnarray}
In this work, we will be interested in the  gauge boson-polarization density matrix elements (DMEs)
defined in a helicity basis as
\begin{eqnarray}
    \rho^{\lambda \lambda'} = \frac{W^{\mu \nu} \epsilon_\mu^{(\lambda)} \epsilon_\nu^{(\lambda')*}}{2 W_T + W_L} ,
\end{eqnarray}
where the gauge boson polarization vectors	
$\epsilon^{(\lambda)}_\mu(q)$ are given by
\begin{eqnarray}
    \epsilon^{(\pm)}_\mu(q) &=& \frac{\mp X^\mu - iY^\mu}{\sqrt{2}} \,, \quad 
\epsilon^{(0)}_\mu(q) =Z^\mu \,.
\end{eqnarray}
Analogously, we denote by
\begin{eqnarray}\label{eq:structure_func_def}
    \rho_i \equiv \frac{W_i}{2W_T + W_L} , i \in \{T,L,\Delta\Delta, \Delta \Delta_p, T_p, \Delta, \nabla, \Delta_p, \nabla_p \},
\end{eqnarray}
the following combinations of DMEs:
\begin{equation} 
    \begin{split}
\rho_{T} &=\,  \frac{1}{2} (\rho^{++}+\rho^{--}),  \\
\rho_{L} &= \, \rho^{00},\ \\
\rho_{\Delta\Delta}  &=  \, \frac{1}{2} (\rho^{-+}+\rho^{+-}), \\
\rho_{\Delta\Delta_{p}} &= \, \frac{i}{2} \left(\rho^{+-}-\rho^{-+}\right), \\
\rho_{T_{p}} &=  \, \frac{1}{2} \left(\rho^{++}-\rho^{--}\right), \\
\rho_{\Delta} &=  \, -\frac{1}{2\sqrt{2}}\left(\rho^{-0}+\rho^{0-}-\rho^{+0}-\rho^{0+}\right), \\
\rho_{\nabla} &= \, \frac{i}{2\sqrt{2}}\left(\rho^{-0}-\rho^{0-}+\rho^{0+}-\rho^{+0}\right), \\
\rho_{\Delta_{p}} &= \, \frac{i}{2\sqrt{2}}\left(-\rho^{0-}+\rho^{-0}-\rho^{0+}+\rho^{+0}\right), \\
\rho_{\nabla_{p}} &= \, -\frac{1}{2\sqrt{2}}\left(\rho^{0-}+\rho^{-0}+\rho^{0+}+\rho^{+0}\right)\, .
\end{split}
\label{eq:cartesian_vs_helicity}
\end{equation}
For completeness, let us give the full expression for the angular distribution in terms of DMEs, which reads
\begin{eqnarray}
    \frac{d N}{d \Omega} &=&  \frac{3}{8\pi}
    \Bigg[ g_T \rho_T + g_L \rho_L + g_\Delta \rho_\Delta  + g_{\Delta\Delta} \rho_{\Delta\Delta} \nonumber \\ 
    & & + c_G g_{T_P} \rho_{T_P} + c_G g_{\nabla_P} \rho_{\nabla_P} + c_G g_\nabla \rho_\nabla + g_{\Delta\Delta_P} \rho_{\Delta\Delta_P} + g_{\Delta_P} \rho_{\Delta_P}
    \Bigg]\, ,
    \label{eq:angular_distribution}
\end{eqnarray}
where $g_i = g_i(\theta,\phi)$ are the angular coefficients 
\begin{eqnarray}
    g_T = 1 + \cos^2\theta \, , \quad & g_L = 1 - \cos^2\theta \, , \quad & g_{T_P} = 2 \cos\theta \, , \nonumber\\ 
    g_{\Delta\Delta} = \sin^2\theta\cos 2\phi \, , \quad & g_\Delta = \sin 2\theta\cos\phi \, , \quad & g_{\nabla_P} = 2\sin\theta\cos\phi \, , \nonumber\\
    g_{\Delta\Delta_P} = \sin^2\theta\sin 2\phi \, , \quad & g_{\Delta_P} = \sin 2\theta\sin\phi \, , \quad & g_{\nabla} = 2 \sin\theta\sin\phi \, . 
\end{eqnarray}
In experimental analyses, one often represents the dilepton angular distribution in terms of angular coefficients $A_i, i = 0, \dots, 7$
\begin{eqnarray}\label{eq:ang-dist-A}
    \frac{\dd N}{\dd\Omega} &=& \frac{3}{16\pi}\Bigg(1 + \cos^2\theta +\frac{A_0}{2}\left(1-3\cos^2\theta \right) + A_1 \sin 2\theta\cos\phi + \frac{A_2}{2}\sin^2\theta\cos 2\phi
    \nonumber \\ & + &
    A_3 \sin\theta\cos\phi + A_4 \cos\theta + A_5 \sin^2\theta\sin 2\phi + A_6 \sin 2\theta\sin\phi + A_7 \sin\theta\sin\phi \Bigg)\, . \nonumber \\
\end{eqnarray}
As the trace of the density matrix equals unity, $2 \rho_T + \rho_L = 1$, one of the DMEs, for example $\rho_T$, can be eliminated, and the remaining ones are easily related to the angular coefficients coefficients $A_i$.

Some aspects are important to be emphasized. First,  the angular coefficients $\rho_i$ depend on the invariant mass, transverse momentum, and
rapidity of the lepton pair, being sensitive to the description of the QCD dynamics.  Second, for the 
purely electromagnetic Drell--Yan process, only the four structure
functions $W_T, \, W_L, \, W_{\Delta}$ and $W_{\Delta \Delta}$ contribute. Third, the parton model predicts \cite{Lam:1978pu} that the quantity
\begin{eqnarray}
    A_{LT} \equiv A_0 - A_2 = \frac{2W_L - 4W_{\Delta \Delta}}{2W_T + W_L} = 2 (\rho_L - 2 \rho_{\Delta \Delta}),
\end{eqnarray}
denoted Lam--Tung relation, is equal to zero. Recent high-statistics measurements of the lepton angular distribution coefficients in $Z$ boson production  in $pp$ collision at the LHC \cite{ATLAS:2016rnf,CMS:2015cyj,LHCb:2022tbc}, over a broad range of the gauge boson transverse momentum,  by the ATLAS, CMS and LHCb experiments revealed a clear violation of the Lam--Tung relation, in contrast with the experimental results at smaller center-of-mass energies. One possible  physical explanation for this
violation  is that QCD radiative effects, which arise at next-to-leading order  (or beyond), become important at LHC energies \cite{Gauld:2017tww}, this effect is particularly obvious in $k_T$--factorization \cite{Motyka:2016lta}. Another possibility is that intrinsic transverse
momenta of the initial partons \cite{Peng:2015spa} is sufficing to generate the Lam--Tung violation. Higher-twist processes involving directly the parton--wave function of the incoming particle may be of relevance in the forward region, but have been studied only in pion--proton and pion--nucleus collisions
\cite{Brandenburg:1994wf,Eskola:1994py}. Other higher twist effects which can be accommodated by our framework are the ones related to non-linear QCD effects \cite{Gelis:2006hy}.
Such aspects motivate the analysis of the angular distributions using the hybrid factorization formalism, which is expected to take into account of the contributions associated with the intrinsic momentum of the target, part of the NLO corrections as well the higher-twist contributions associated with the non-linear QCD effects.

We now go on to calculate the DMEs for the process of gauge boson emission from a projectile (anti--)quark scattering of the gluon field of the second hadron. It will turn out that, at the lowest order considered by us, only the first six terms in the angular distribution Eq.(\ref{eq:angular_distribution}) contribute.
The out-of-plane correlations $\propto \sin \phi, \sin 2 \phi$ are not generated.
Notice that these contributions change sign when going from the right--handed to the left--handed coordinate system, signalling the violation of parity symmetry.
A careful analysis of discrete symmetries \cite{Martens:2017cvj} shows that the relevant structure functions are also odd under time--reversal. Such contributions, for unpolarized partons, can only be due to imaginary parts generated by loop--corrections \cite{Hagiwara:1984hi}. We therefore can anticipate their vanishing in our calculation presented below.
\section{Helicity density matrix elements from the $q \to Gq$ process in the color--dipole S--matrix framework}
\label{sec:DMEs}
The previous equations indicate that the helicity structure functions can be obtained from the different contractions $\epsilon_{\mu}^{(\lambda)}\epsilon_{\nu}^{(\lambda')*} W^{\mu \nu}$. Therefore in this section we focus on the helicity density matrix for the gauge-boson production. 
In the following, we will further simplify the calculation by replacing the Breit--Wigner distribution of the dilepton mass by a delta function. We therefore do not discuss the dependence on lepton invariant mass, nor do we include the $Z^0-\gamma$ interference. 
We start from the cross section for inclusive gauge--boson production which can be expressed through the hadronic tensor as
\begin{eqnarray}
\frac{d \sigma (pp \to G(x_F,\qv) X)}{dx_F d^2 \qv} = \frac{1}{32 \pi^3 s x_F} ( 2 W_T + W_L) \, .
\end{eqnarray}
We are interested in the forward rapidity region of one of the protons which has, say, a large lightcone plus-momentum, and the gauge boson carries a fraction $x_F$ thereof. By $\qv$ we denote the transverse momentum of the gauge boson in the $pp$ center-of-mass frame.
We will also below use the notation $q_T = |\qv|$ for its absolute value.

Now the polarization DMEs in the helicity basis are related to the hadronic tensor via  
\begin{eqnarray}
\rho^{\lambda \lambda'} \, \frac{d \sigma (pp \to G(x_F,\qv) X)}{dx_F d^2 \qv} = \frac{1}{32 \pi^3 s x_F} \, 
\epsilon_{\mu}^{(\lambda)}\epsilon_{\nu}^{(\lambda')*} W^{\mu \nu} \,\,.
 \label{eq:cross_sec_Wmunu}
\end{eqnarray}
As we are interested in the behavior of these quantities at forward rapidities, we use a hybrid factorization formula for the lhs. of Eq.~\eqref{eq:cross_sec_Wmunu}, which implies that
\begin{eqnarray}
    \rho_{\lambda \lambda'}\frac{d \sigma\big(pp\rightarrow G(x_F,\qv) X\big)}{\dd x_F \dd^{2}\qv} &=& \sum_f \int \dd x_1 \dd z \,\delta(x_F - zx_1)
   \Bigg[ q_f(x_1,\mu^2)+\overline q_f(x_1,\mu^2) \Bigg] 
   \nonumber \\ && \, \, \times 
   \hat\rho_{\lambda \lambda'}\frac{\dd\hat\sigma\big(qp\rightarrow G(z,\qv)X\big)}{\dd z \dd^{2}\qv}\,, 
   \label{Eq:bosonpp}
\end{eqnarray}
where $q_f(x_1,\mu^2)$ are the projectile parton distributions with a momentum fraction $x_1$ at a factorization scale $\mu^2 \approx \qv^2 + M_G^2$, $z$ is the fraction of the quark's longitudinal momentum carried by the gauge boson. Moreover, the partonic density matrix is given in the color-dipole $S$-matrix  approach by
\begin{equation}
    \begin{split}
        \hat \rho^{\lambda\lambda'}\frac{\dd\sigma(qp\rightarrow GX)}{\dd z\dd^{2}\qv}=\  & \frac{1}{2(2\pi)^{2}}\overline{\sum_{\eta\eta'}}\int\dd^{2}\rr\dd^{2}\rr'\!e^{-i \qv(\rr-\rr')}\Psi_{\eta\eta'}^{(\lambda)}(z,\rr)\Psi_{\eta\eta'}^{(\lambda')\dagger}(z,\rr') \\
 & \times\Big[\sigma(x,z\rr)+\sigma(x,z\rr')-\sigma(x,z(\rr-\rr'))\Big],
    \end{split}
    \label{Eq:parton_spectraIP}
\end{equation}
where $\Psi_{\eta\eta'}^{(\lambda)}$ is the light front wave function (LFWF) for the $q_{\eta}\rightarrow Gq'_{\eta'}$
transition in the impact parameter space, with $\eta,\eta',\lambda,\lambda'$ denoting the helicities of particles, and $\sigma(x,\rr)$ is the color dipole--proton cross section \cite{Nikolaev:1990ja}, which is determined by the QCD dynamics.
For an early discussion of Drell--Yan lepton angular distributions in the color-dipole approach, see \cite{Brodsky:1996nj}. 
The fact that the gauge boson at the leading order does not interact with the color--field is a major simplification compared to the production of the gauge boson of the strong interactions -- the gluon, see \cite{Nikolaev:2005dd,Nikolaev:2005zj,Nikolaev:2005qs}.

In what follows, we will use the results for the LFWF obtained in Ref.\cite{Bandeira:2024zjl} in order to derive the elements of the partonic density matrix. As demonstrated in that reference, the calculation can be performed in two equivalent representations, the impact parameter and momentum space representations, which are related by a Fourier transform. Here we will perform our study in the momentum space representation. In order to do that, we will use in Eq. (\ref{Eq:parton_spectraIP})  the relation between the dipole--proton cross section and the unintegrated gluon distribution, $f(x,\boldsymbol{k})$, given by
\begin{equation}
\sigma(x,\rr)=\frac{1}{2}\int\dd^{2}\boldsymbol{k}f(x,\boldsymbol{k})\left(1-e^{-i \boldsymbol{kr}}\right)\left(1-e^{i \boldsymbol{kr}}\right).
\end{equation}
Inserting this expression into Eq.(\ref{Eq:parton_spectraIP}), we obtain the following representation, now fully in momentum space:
\begin{eqnarray}
\hat \rho^{\lambda\lambda'}\frac{d\sigma(qp\rightarrow GX)}{d z d^{2}\qv} &=&   \frac{1}{2(2\pi)^{2}}\overline{\sum_{\eta\eta'}}\int d^{2}\boldsymbol{k}f(x,\boldsymbol{k}) \ \nonumber \\
&& \times \left(\Psi_{\eta' \eta}^{(\lambda)}(z,\qv)-\Psi_{\eta' \eta}^{(\lambda)}(z,\qv-z\boldsymbol{k})\right)\left(\Psi_{\eta'\eta}^{(\lambda')}(z,\qv)-\Psi_{\eta'\eta}^{(\lambda')}(z,\qv-z\boldsymbol{k})\right)^{\dagger} \nonumber \\
&=&  \frac{1}{2(2\pi)^{2}} \int \, d^2 \kv \, f(x_2,\kv) \, I^{(\lambda, \lambda')}(z,\qv,z\kv) \, . 
\label{eq:parton_level_xsec}
\end{eqnarray}
Here, the argument of the unintegrated gluon distribution is calculated from the transverse mass of the $q+G$ system:
\begin{eqnarray}
    x_1 x_2 s = M^2_{qG} + \kv^2 = \frac{M_G^2 + \qv^2}{z} + \frac{m_f^2 + (\kv - \qv)^2}{1-z} \, .
\end{eqnarray}
It is convenient to write the combinations of LFWFs relevant for the density matrix in the form
\begin{eqnarray}
\Psi^{(\lambda)}_{\eta' \eta}(z,\qv) - \Psi^{(\lambda)}_{\eta' \eta}(z, \qv - z \kv) = \, 
\chi^\dagger_{\eta'} {\cal O}^{(\lambda)} \, \chi_\eta \, , 
\end{eqnarray}
so that the relevant integrands of Eq. (\ref{eq:parton_level_xsec}) can be written as
\begin{eqnarray}
    I^{(\lambda, \lambda')}(z,\qv,z\kv) = 
    \frac{1}{2} 
    \mathrm{Tr}[{\cal O}^{(\lambda)}
    {\cal O}^{(\lambda')\dagger}] \, . 
    \label{eq:traces}
\end{eqnarray}
Let us now evaluate these traces, first constructing operators ${\cal O}^{(\lambda)}$ from the explicit LFWF given in appendix \ref{app:LFWF}. Using the shorthand notations
\begin{align}
\Phi_{0}=\  & \frac{1}{\qv^{2}+\epsilon^{2}}-\frac{1}{(\qv-z\kv)^{2}+\epsilon^{2}},
\quad \bPhi= \frac{\qv}{\qv^{2}+\epsilon^{2}}-\frac{\qv - z \kv}{(\qv- z \kv)^{2}+\epsilon^{2}},\label{eq:Phi}
\end{align}
where $\epsilon^2 = (1-z) M^2 + z(m_b^2 -m_a^2) + z^2 m_a^2$,
and
\begin{align*}
\Gamma_{V}\equiv\  & m_{b}-(1-z)m_{a}, \quad 
\Gamma_{A}\equiv  m_{b}+(1-z)m_{a}\\
\Lambda_{V}\equiv\  & z^{2}m_{a}\left(m_{b}-m_{a}\right)-z\left(m_{b}^{2}-m_{a}^{2}\right)-2(1-z)M^{2}\\
\Lambda_{A}\equiv\  & z^{2}m_{a}\left(m_{b}+m_{a}\right)+z\left(m_{b}^{2}-m_{a}^{2}\right)+2(1-z)M^{2}.
\end{align*}
we obtain
\begin{eqnarray}
{\cal O}^{(\pm)}  = C_f^G  \sqrt{z}  &&\Big\{ \frac{2-z}{z} \, (\bPhi \cdot \be^*(\pm) ) \Big( g^G_{V,f} \openone + g^G_{A,f} \sigma_3 \Big) 
+ i (\bPhi \times \be^*(\pm)) \cdot \be_z \Big( g^G_{V,f} \sigma_3 + g^G_{A,f} \openone \Big) \nonumber \\
&& - \Phi_0 \, (\bsigma \cdot \be^*(\pm)) 
\Big( g^G_{V,f} \Gamma_V \sigma_3 + g^G_{A,f} \Gamma_A \openone \Big) \Big\}  \, ,
\end{eqnarray}
and 
\begin{eqnarray}
{\cal O}^{(0)} =\frac{C_f^G}{\sqrt{z}M} \Big\{ \Phi_0 \Big( g^G_{V,f} \Lambda_V \openone - g^G_{A,f} \Lambda_A \sigma_3 \Big) + z (\bsigma \cdot \bPhi) \Big( g^G_{V,f} (m_b - m_a) \sigma_3 - g^G_{A,f}(m_b + m_a) \openone \Big) \Big\} \, . \nonumber \\
\end{eqnarray}
Now, the traces of Eq.~\eqref{eq:traces} are readily evaluated using
\begin{eqnarray}
\mathrm{Tr}\left(\openone\right)= 2, \quad
\mathrm{Tr}\left(\sigma_{i}\right)= 0, \quad
\mathrm{Tr}\left(\sigma_{i}\sigma_{j}\right)= 2\delta_{ij}, \quad  
\mathrm{Tr}\left(\sigma_{l}\sigma_{j}\sigma_{k}\right)= 2i \, \varepsilon_{ljk}.
\end{eqnarray}
Consequently, for the transverse polarizations of gauge bosons, one has:
\begin{eqnarray}
 I^{(\pm, \pm)}(z,\qv,z\kv) &=&  |C_f^G|^2 \Big\{  \Big( (g_{V,f}^G)^2 + (g_{A,f}^G)^2 \Big) \, \frac{1 + (1-z)^2}{z} \, \bPhi^2  \pm 2 g_{V,f}^Gg_{A,f}^G(2-z)\bPhi^2 \nonumber \\
 &&+ \Big( (g_{V,f}^G)^2 \Gamma_V^2 + (g_{A,f}^G)^2 \Gamma_A^2 \pm 2 g^G_{V,f} g^G_{A,f} \Gamma_V \Gamma_A \Big) z \Phi_0^2 \Big\} \,  \\
 I^{(\pm, \mp)}(z,\qv,z \kv) &=& |C_f^G|^2 \, 
 \Big(  (g_{V,f}^G)^2 + (g_{A,f}^G)^2 \Big) 2 \, \frac{1-z}{z} \Big( \Phi_y^2 - \Phi_x^2 \Big) \nonumber \\
 &=&  |C_f^G|^2 \, 
 \Big(  (g_{V,f}^G)^2 + (g_{A,f}^G)^2 \Big) 2 \, \frac{1-z}{z} \Big\{ \bPhi^2 - 2 \Big( \frac{\qv \cdot \bPhi}{|\qv|} \Big)^2 \Big\} \, , 
\end{eqnarray}
while for the longitudinally polarized bosons this results in
\begin{eqnarray}
I^{(0,0)}(z,\qv,z\kv) &=&  \frac{|C_f^G|^2}{zM^2} \Big\{ \Big( (g_{V,f}^G)^2 \Lambda_V^2 + (g_{A,f}^G)^2 \Lambda_A^2 \Big) \Phi_0^2 \nonumber \\
&&+ z^2 \Big( (g_{V,f}^G)^2 (m_b - m_a)^2 + (g_{A,f}^G)^2 (m_b + m_a)^2 \Big) \bPhi^2  \Big\}\, . 
\end{eqnarray}
Finally, the interference contributions are given by
\begin{eqnarray}
I^{(\pm,0)}(z,\qv,z\kv) &=& \frac{|C_f^G|^2}{M \sqrt{2}} \Big\{ \mp \frac{2-z}{z^2} \Big( (g_{V,f}^G)^2 \Lambda_V - 
(g_{A,f}^G)^2 \Lambda_A \Big) 
- \frac{g_{V,f}^G g^G_{A,f}}{z} ( \Lambda_V - \Lambda_A ) \nonumber \\
&&\pm 
 (g_{V,f}^G)^2 \Gamma_V (m_b - m_a) \mp (g_{A,f}^G)^2 \Gamma_A (m_b + m_a) \nonumber \\&& + g_{V,f}^G g_{A,f}^G \Big( \Gamma_A (m_b - m_a) - \Gamma_V ( m_b + m_a)  \Big) 
\Big\} z \Phi_0 \, \frac{\qv \cdot \bPhi}{|\qv|} \\
I^{(0,\pm)}(z,\qv,z\kv) &=& \Big(I^{(\pm,0)}(z,\qv,z\kv) \Big)^* = I^{(\pm,0)}(z,\qv,z\kv) \,\,.
\label{eq:LT_identity}
\end{eqnarray}
The above expressions can be used to calculate the different elements of polarization density matrix, $\rho^{\lambda \lambda'}$. Moreover, these quantities  allow us to derive the DMEs defined in Eq.~\eqref{eq:cartesian_vs_helicity}. 
We start from the parity even transverse structure functions
\begin{eqnarray}
    I_T(z,\qv, z\kv) =  |C_f^G|^2 \Big\{  \Big( (g_{V,f}^G)^2 + (g_{A,f}^G)^2 \Big) \, \frac{1 + (1-z)^2}{z} \, \bPhi^2  + \Big( (g_{V,f}^G)^2 \Gamma_V^2 + (g_{A,f}^G)^2 \Gamma_A^2 \Big) z \Phi_0^2 \Big\} \nonumber \\
\end{eqnarray}
and
\begin{eqnarray}
    I_{\Delta \Delta}(z,\qv, z \kv) = |C_f^G|^2 \, 
 \Big(  (g_{V,f}^G)^2 + (g_{A,f}^G)^2 \Big) 2 \, \frac{1-z}{z} \Big\{ \bPhi^2 - 2 \Big( \frac{\qv \cdot \bPhi}{|\qv|} \Big)^2 \Big\} \, , 
\end{eqnarray}
There is only one nonvanishing parity odd transverse combination, namely
\begin{eqnarray}
I_{T_p} (z,\qv,z\kv) = |C_f^G|^2 \, 2 g^G_{V,f} g^G_{A,f} \Big\{ (2-z) \bPhi^2 + z \Phi_0^2 \Big\} ,
\end{eqnarray}
while $I_{\Delta \Delta_p} \equiv 0$. For the longitudinal polarizations we have
\begin{eqnarray}
   I_L(z,\qv,z\kv) &=& I^{(0,0)}(z,\qv,z\kv) \, \nonumber \\
   &=& \frac{|C_f^G|^2}{zM^2} \Big\{ \Big( (g_{V,f}^G)^2 \Lambda_V^2 + (g_{A,f}^G)^2 \Lambda_A^2 \Big) \Phi_0^2 \nonumber \\
&&+ z^2 \Big( (g_{V,f}^G)^2 (m_b - m_a)^2 + (g_{A,f}^G)^2 (m_b + m_a)^2 \Big) \bPhi^2  \Big\}\, . 
  \end{eqnarray}
As far as the longitudinal-transverse interference structure functions are concerned, the identities of Eq. (\ref{eq:LT_identity}) immediately yield the vanishing of%
\begin{eqnarray}
 I_\nabla = I_{\Delta_p} \equiv 0 \, .   
\end{eqnarray}
We obtain one nonvanishing parity even LT-interference structure function with the kernel
\begin{eqnarray}
    I_{\Delta}(z,\qv,z\kv)  &=& \frac{|C_f^G|^2}{M} \Big\{ \frac{2-z}{z} \Big( (g^G_{A,f})^2 \Lambda_A - (g^G_{V,f})^2 \Lambda_V \Big)  \nonumber \\
    &&+ z \Big( (g^G_{V,f})^2 \Gamma_V (m_b - m_a) - (g^G_{A,f})^2 \Gamma_A(m_b + m_a) \Big) \Big \}   \, \Phi_0 \frac{\qv \cdot \bPhi}{|\qv|}, 
\end{eqnarray}
and a parity odd one;
\begin{eqnarray}
    I_{\nabla_p}(z,\qv,z\kv) = \frac{|C_f^G|^2}{M} \, g^G_{V,f} g^G_{A,f} \Big\{ \Lambda_V - \Lambda_A + z \Big( (m_b - m_a) \Gamma_A - (m_b + m_a) \Gamma_V\Big) \Big\} \, \Phi_0 \frac{\qv \cdot \bPhi}{|\qv|} . \nonumber \\
\end{eqnarray}
For the case of virtual photons, our results are in agreement with the results of \cite{Schafer:2016qmk}. Notice that for massless quarks $\Gamma_{V,A} \to 0$, and the parity- conserving structure functions of electroweak gauge bosons become proportional to the ones for virtual photons. For finite quark masses this is evidently not the case, due to the non-conservation of the axial current.
Substantial simplifications occur in the limit of massless quarks, which is relevant for the practical applications.
In particular, in the massless quark limit, parity even structure functions will be proportional to $(g_{V,f}^G)^2 + (g_{A,f}^G)^2 $, while for finite masses also the combination $(g_{V,f}^G)^2 - (g_{A,f}^G)^2 $ will enter.
\section{Results}
\label{sec:res}

In this section we will present our predictions for the transverse momentum dependence of the six nonvanishing combinations of helicity DMEs $\rho_i$, with $i = T,L, \Delta, \Delta\Delta, T_p$ and $\nabla_p$. We will consider $pp$ collisions at the LHC energy of $\sqrt{s} = 14$ TeV
and focus on dileptons produced at forward rapidities ($2.0 \le y \le 4.0$), where the hybrid factorization assumed in our calculations is expected to be valid. The results derived in the previous section indicate that structure functions $W_i$ are linearly proportional to the target unintegrated gluon distribution,  $f(x,\boldsymbol{k})$. Together with the collinear parton distributions of quarks and antiquarks, such a quantity is the main input in our calculations and is determined by the QCD dynamics.

\subsection{Collinear quark PDFs and Unintegrated gluon distributions}

At forward rapidities, the collinear distributions $q(x,\mu^2)$ and $\bar{q}(x,\mu^2)$ in Eq.~\eqref{Eq:bosonpp} are probed at large values of $x$, where the parton distribution functions obtained by the distinct groups that perform the global analysis are similar.
In what follows,  we use the CT14LL parameterization
of the CTEQ collaboration \cite{Dulat:2015mca} for the collinear distributions $q(x,\mu^2)$ and $\bar{q}(x,\mu^2)$ in Eq.~\eqref{Eq:bosonpp}. As the angular coefficients are given in terms of ratios of structure functions, the impact of next - to - leading order corrections to the PDFs on our results is expected to be small.


\begin{figure}[t]
    \centering
    \includegraphics[width=0.45\linewidth]{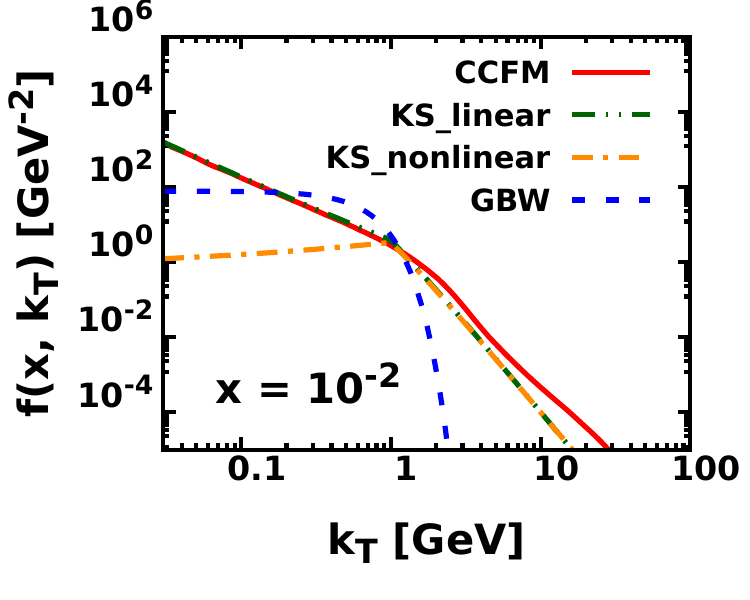}
    \includegraphics[width=0.45\linewidth]{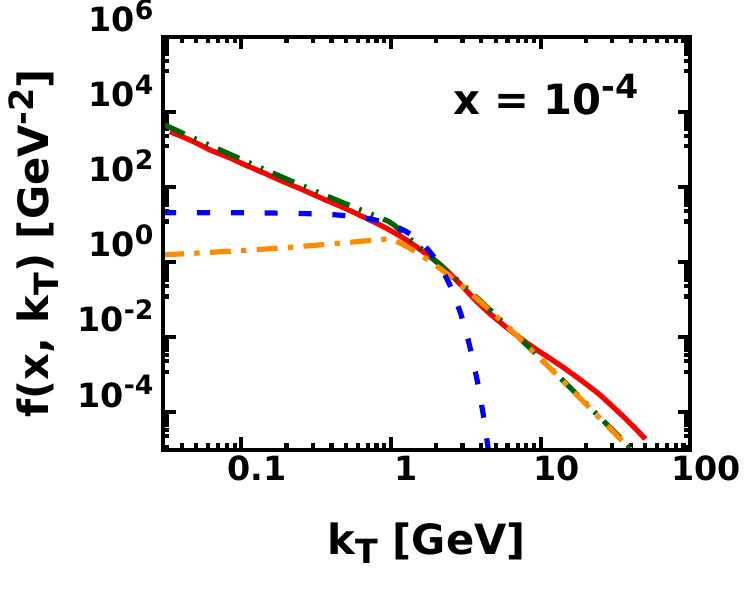} 
    \caption{A set of unintegrated gluon distributions at $x = 10^{-2}$ in left panel and at $x = 10^{-4}$ in the right panel.}
    \label{fig:ugds-mu100}
\end{figure}

On the other hand, as the small - $x$ region is probed in the proton target,  we will consider four distinct models for the proton UGD: two based on a linear QCD dynamics (CCFM and KS--linear), proposed in Refs.~\cite{Hautmann:2013tba,Kutak:2012rf,Golec-Biernat:1998zce}, and two based on a non-linear QCD dynamics (Golec-Biernat--W\"usthoff (GBW) and KS--nonlinear). In particular, the KS--nonlinear UGD is based on the solution of the Balitsky--Kovchegov (BK) equation, with the free parameters adjusted using the $ep$ HERA data, while the KS--linear UGD has been obtained disregarding the non-linear term in the BK equation. As a consequence, the comparison between the predictions derived using these two models allow us to estimate how sensitive is the quantity to non-linear effects in the QCD dynamics. For completeness, we also present the predictions associated with the CCFM and GBW UGDs, which also are able to describe the HERA data at small values of the Bjorken-$x$ variable, but are based on distinct assumptions. In particular, the CCFM parametrization is evaluated assuming $\mu = 100$ GeV. In Fig.~\ref{fig:ugds-mu100} we present a comparison between these distinct UGDs for two different values of $x$. The UDGs differ on the predicted transverse momentum dependence. At small $k_T$ the CCFM and KS--linear models are similar, but become distinct with increasing transverse momentum. In both models, $f$ increases for $k_T\rightarrow 0$.   In contrast, the GBW and KS--nonlinear models predict that $f \rightarrow 0$ when $k_T\rightarrow 0$. Another important aspect is that the value of $k_T$ where the KS--nonlinear and KS--linear UGDs become identical is dependent of $x$, increasing for smaller values of $x$. Such a result is expected, since the transition line between the non-linear and linear regimes of QCD dynamics is determined by the saturation scale $Q_s$, which depends on $x$ (For a more detailed discussion see, e.g., Ref.~\cite{Golec-Biernat:1998zce,Golec-Biernat:1999qor,Bandeira:2022dvu}).


\subsection{Helicity DMEs}




\begin{figure}[t]
    \centering
    \includegraphics[width=0.45\linewidth]{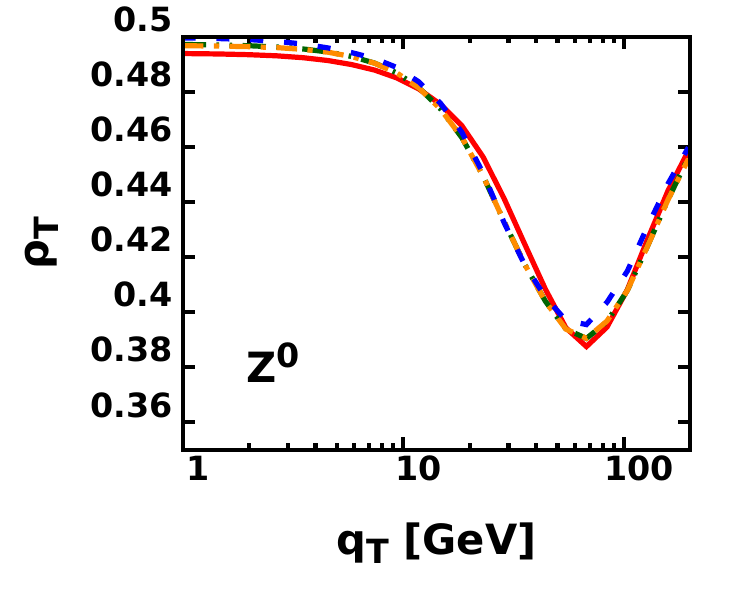}
    \includegraphics[width=0.45\linewidth]{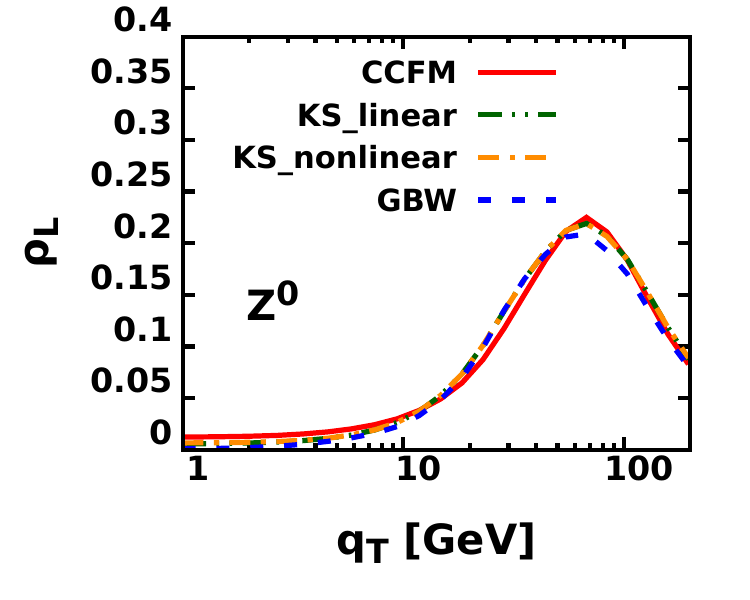} \\
    \includegraphics[width=0.45\linewidth]{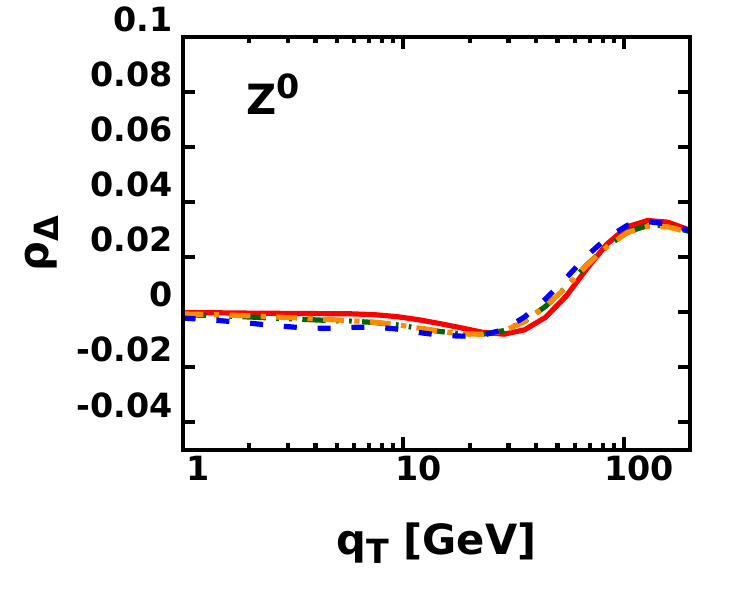} 
    \includegraphics[width=0.45\linewidth]{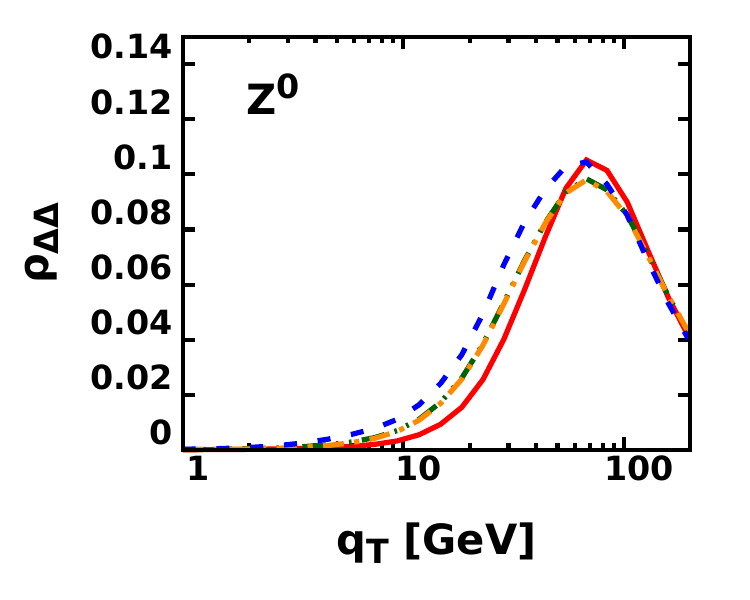}\\
    \includegraphics[width=0.45\linewidth]{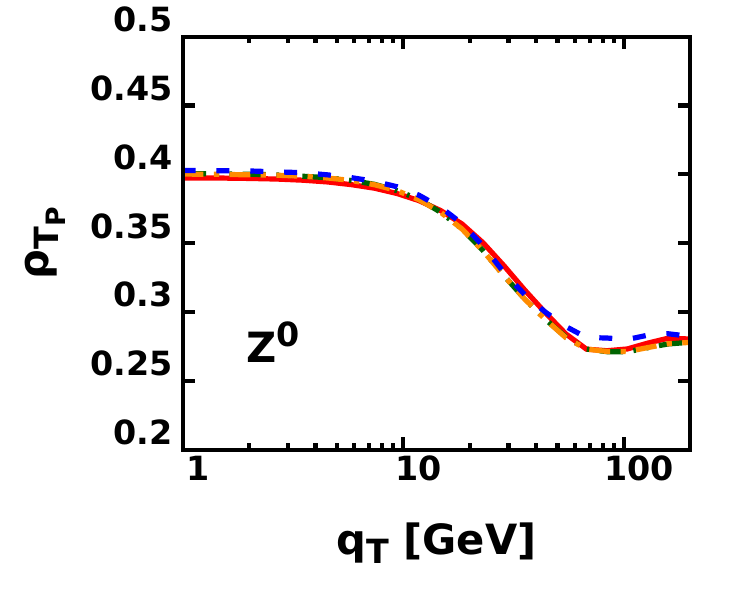}
    \includegraphics[width=0.45\linewidth]{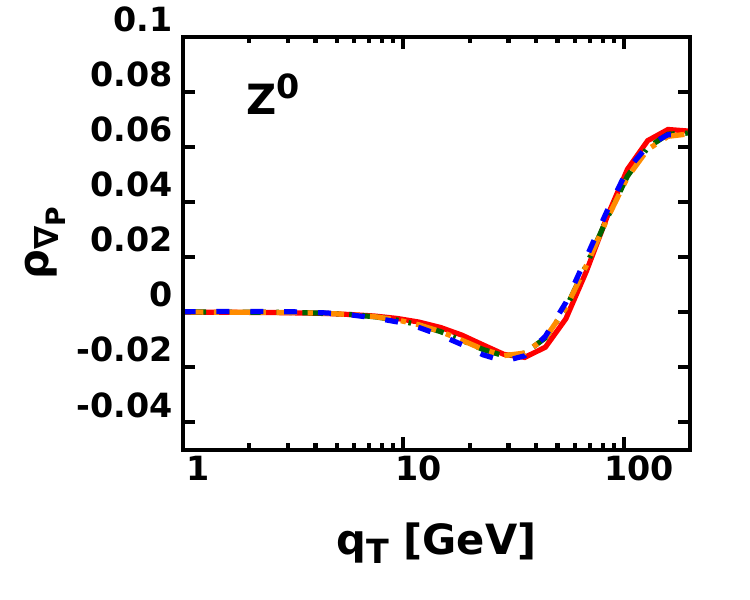}
    \caption{The density matrix elements $\rho_{i = T,\,L,\,\Delta,\,\Delta\Delta,\,T_P,\,\nabla_P}$ for the $Z^0$ gauge boson in the GJ frame, evaluated in the forward rapidity range $2 \le y \le 4$.}
    \label{fig:DME-Z0-GJframe}
\end{figure}

\begin{figure}[h]
    \centering
    \includegraphics[width=0.45\linewidth]{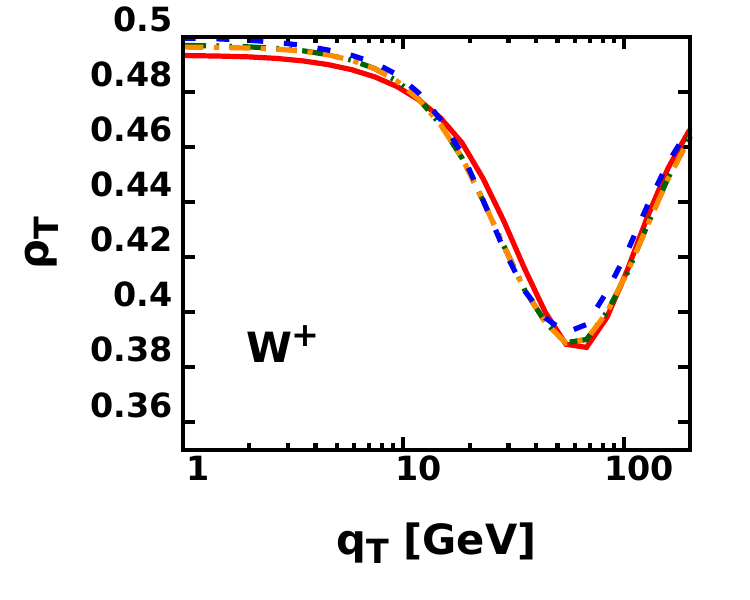}
    \includegraphics[width=0.45\linewidth]{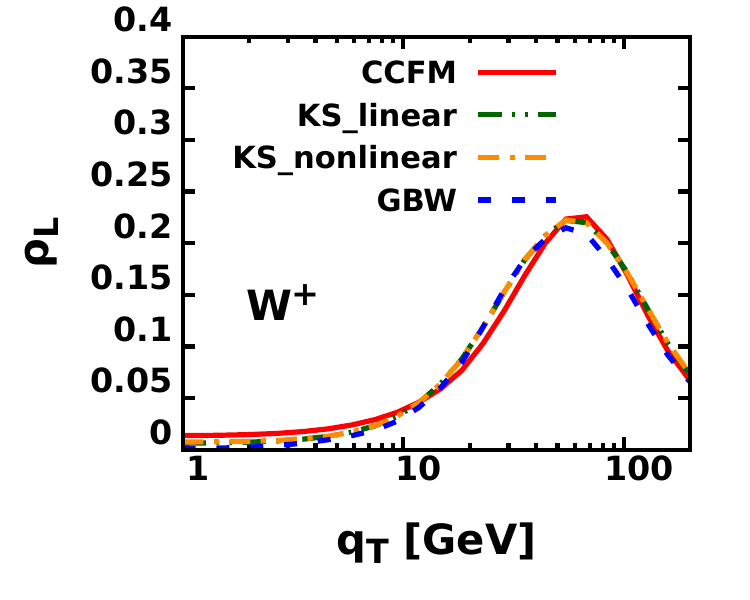} \\
    \includegraphics[width=0.45\linewidth]{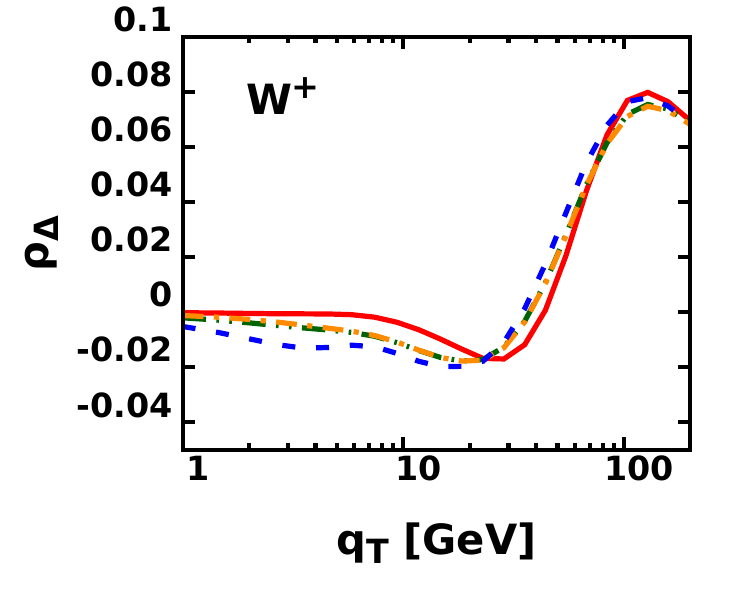} 
    \includegraphics[width=0.45\linewidth]{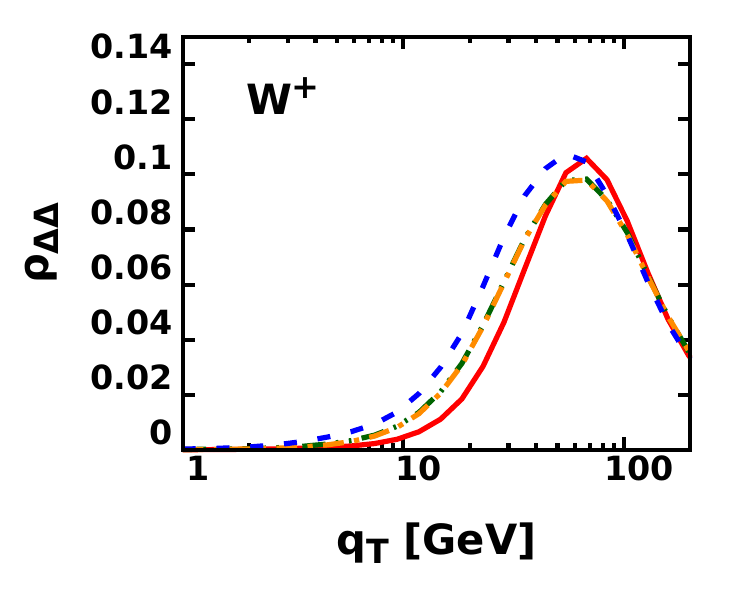} \\
    \includegraphics[width=0.45\linewidth]{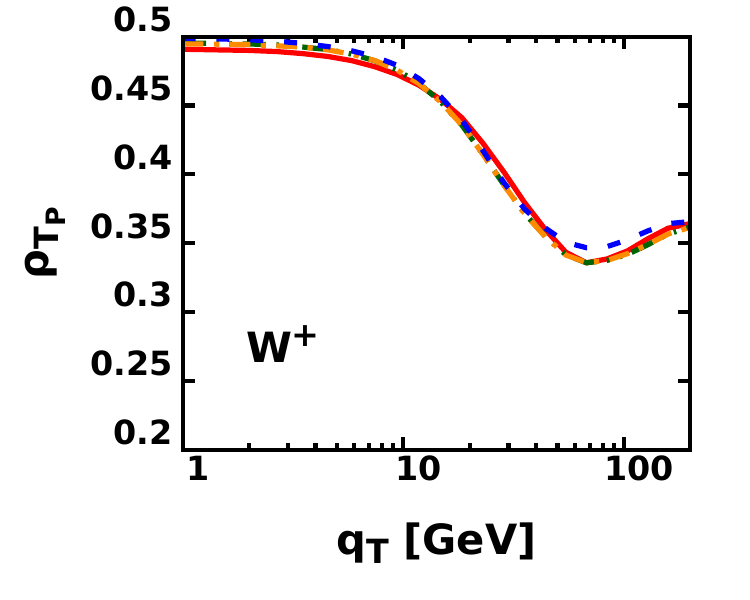} 
    \includegraphics[width=0.45\linewidth]{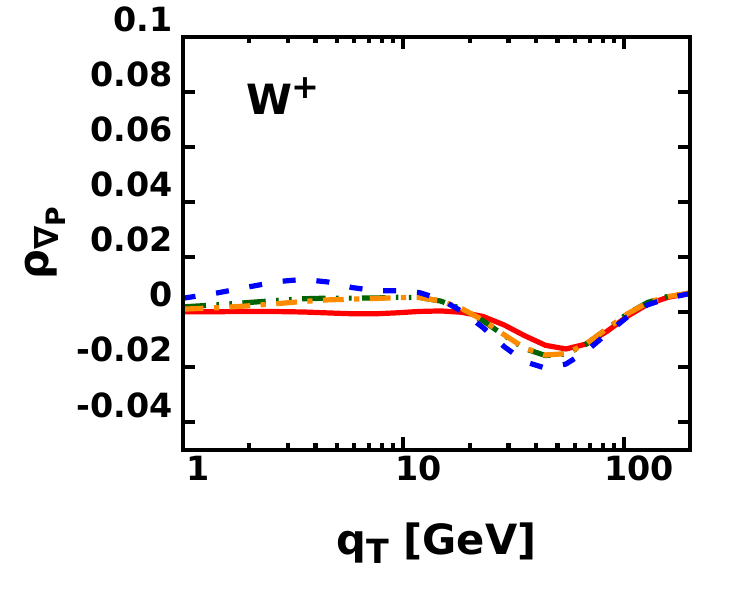} 
    \caption{The density matrix elements $\rho_{i = T,\,L,\,\Delta,\,\Delta\Delta,\,T_P,\,\nabla_P}$ for the $W^+$ gauge boson in the GJ frame, evaluated in the forward rapidity range $2 < y < 4$ }
    \label{fig:DME-Wp-GJframe}
\end{figure}

\begin{figure}[h]
    \centering
    \includegraphics[width=0.45\linewidth]{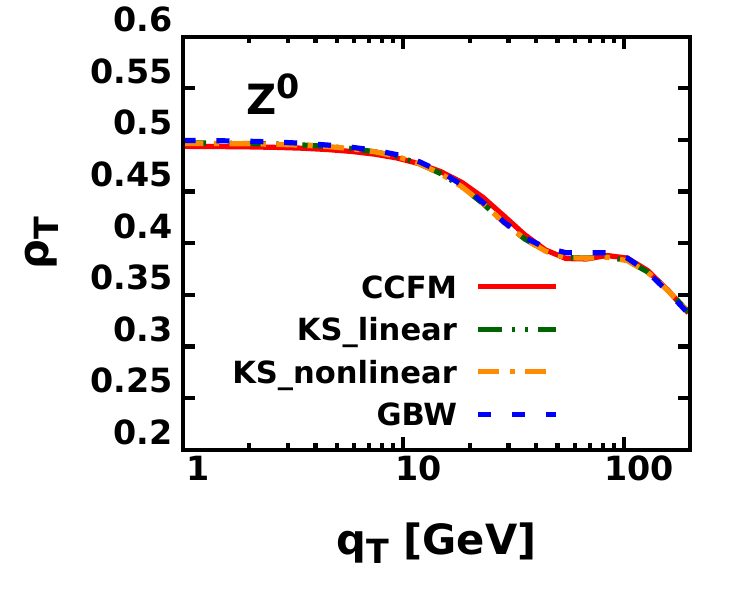}
    \includegraphics[width=0.45\linewidth]{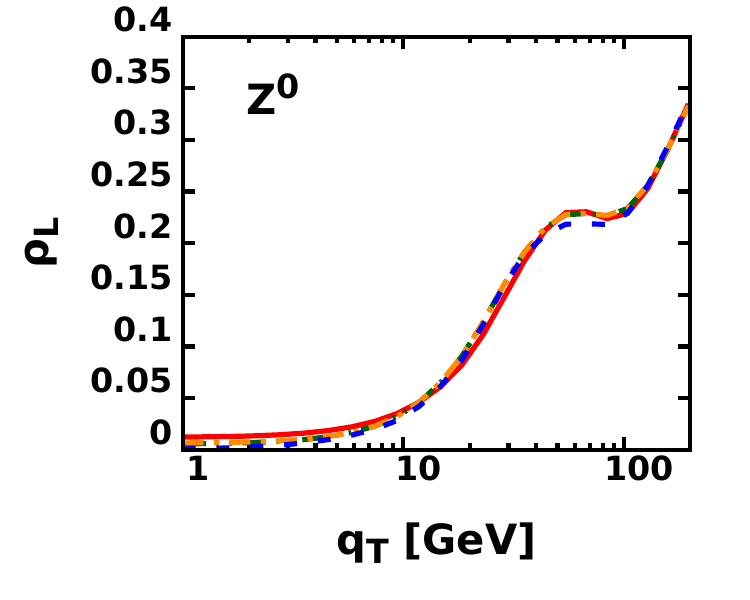} \\
    \includegraphics[width=0.45\linewidth]{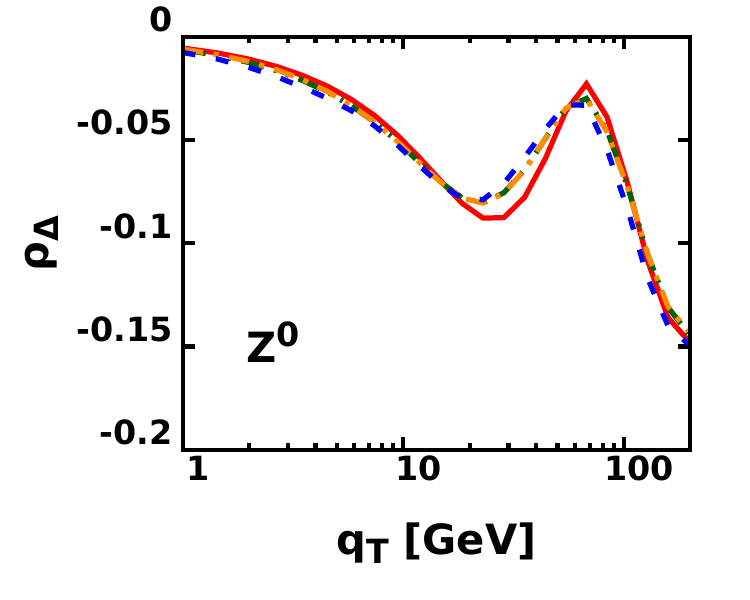} 
    \includegraphics[width=0.45\linewidth]{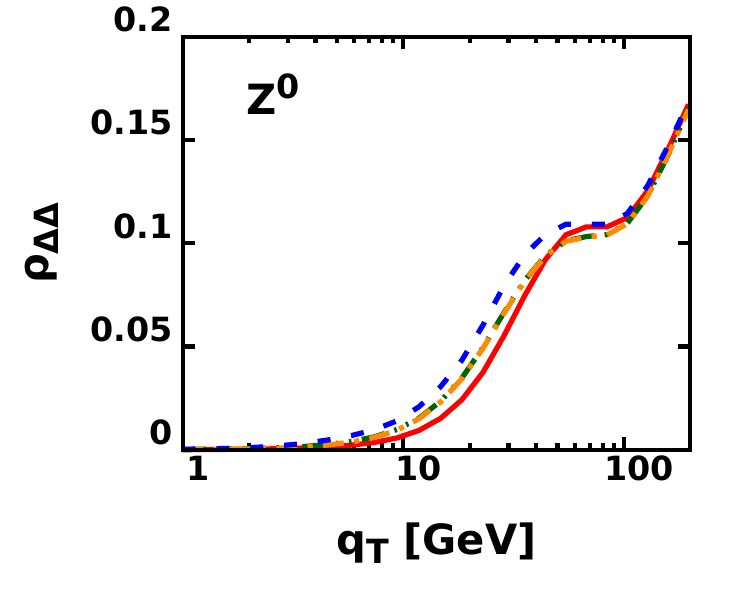} \\
    \includegraphics[width=0.45\linewidth]{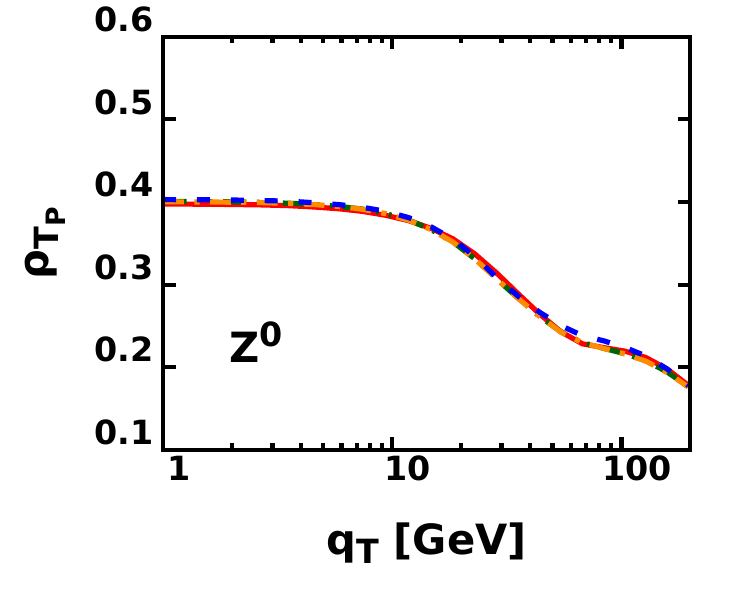}
    \includegraphics[width=0.45\linewidth]{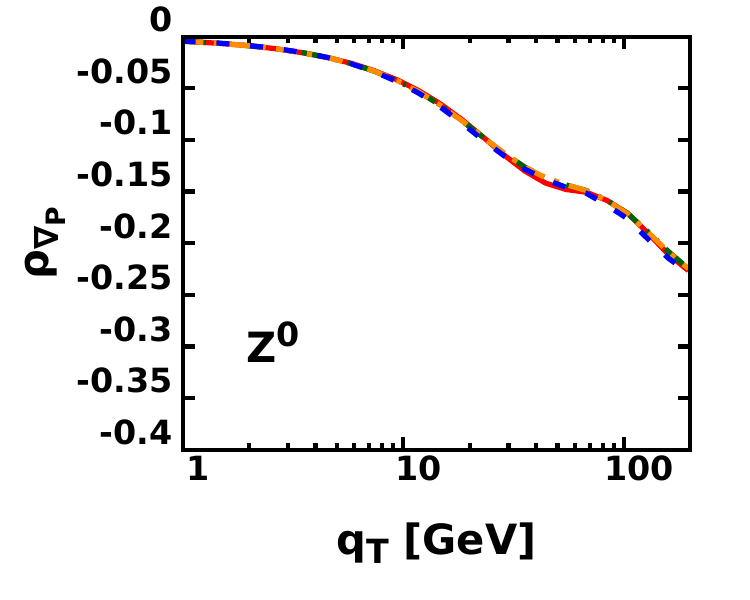} 
    \caption{The density matrix elements $\rho_{i = T,\,L,\,\Delta,\,\Delta\Delta,\,T_P,\,\nabla_P}$ for the $Z^0$ gauge boson in the CS frame, evaluated in the forward rapidity range $2 \le y \le 4$.
    }
    \label{fig:DME-Z0-CSframe}
\end{figure}

\begin{figure}[h]
    \centering
    \includegraphics[width=0.45\linewidth]{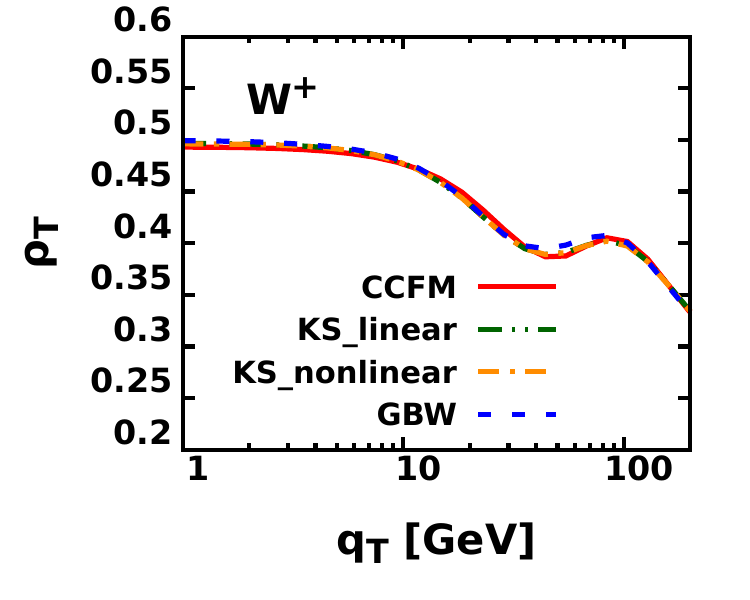}
    \includegraphics[width=0.45\linewidth]{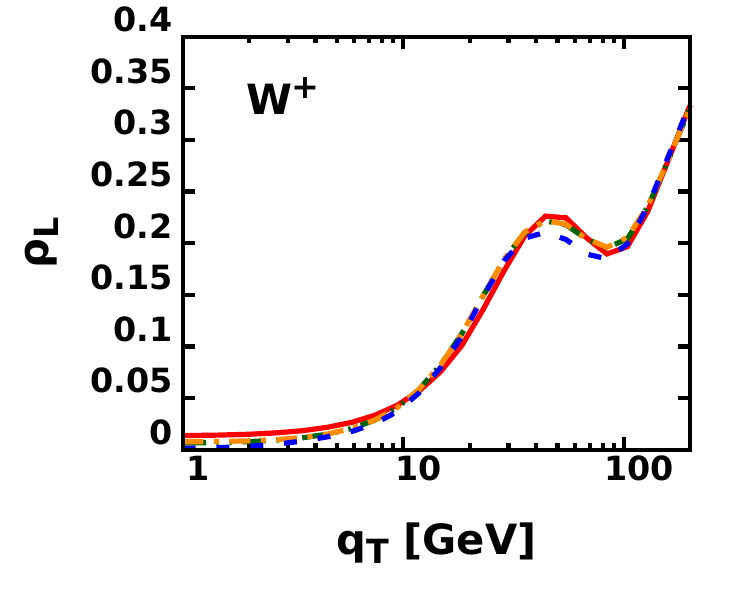} \\
    \includegraphics[width=0.45\linewidth]{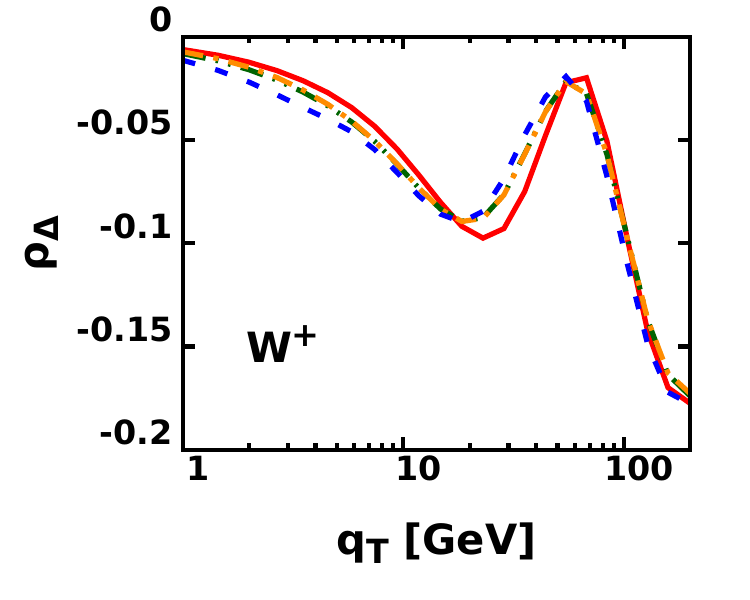} 
    \includegraphics[width=0.45\linewidth]{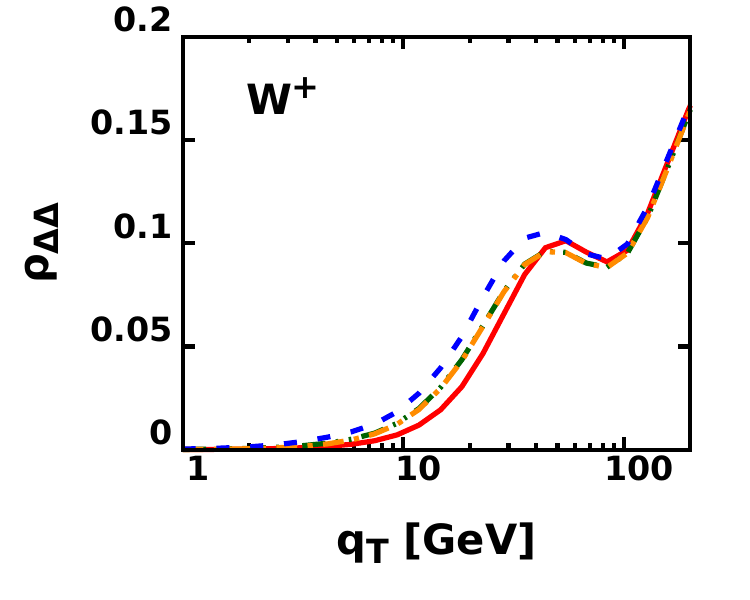} \\
    \includegraphics[width=0.45\linewidth]{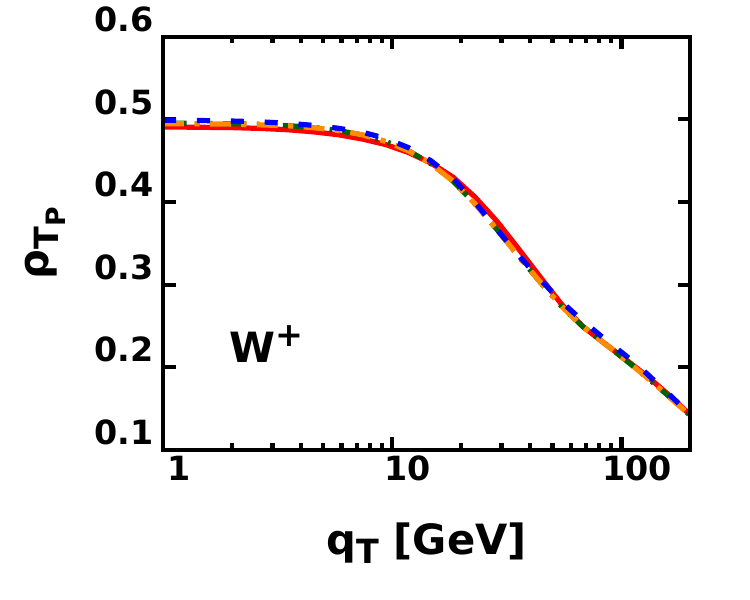} 
    \includegraphics[width=0.45\linewidth]{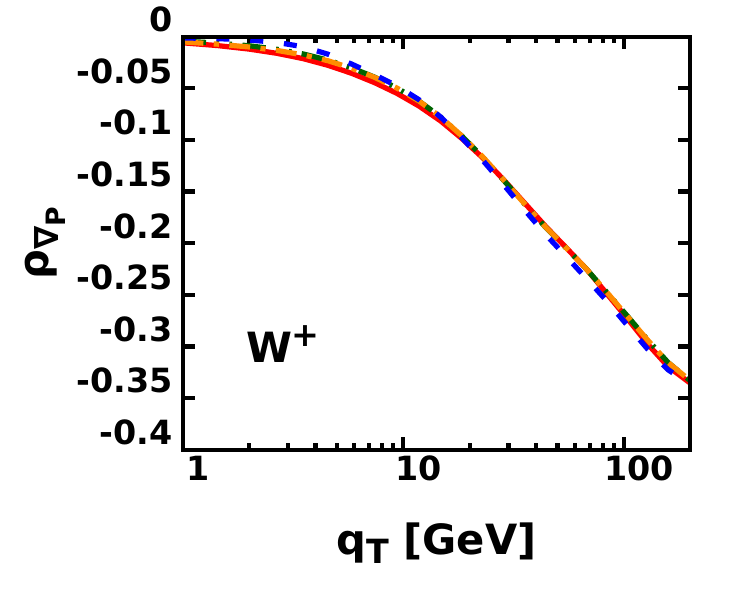}  
    \caption{The density matrix elements $\rho_{i = T,\,L,\,\Delta,\,\Delta\Delta,\,T_P,\,\nabla_P}$ for the $W^+$ gauge boson in the CS frame, evaluated in the forward rapidity range $2 \le y \le 4$.}
    \label{fig:DME-Wp-CSframe}
\end{figure}

In order to present our predictions for the DMEs, we have to specify the choice of the coordinate system in the gauge boson rest frame, with respect to which the angles of decay products are measured. Such choice is arbitrary, and distinct coordinate systems are often used in different studies in the literature  (See the discussion in the textbook \cite{Faccioli:2022peq}). Two of the more frequently used frames are the Gottfried--Jackson (GJ) and Collins--Soper (CS) frames, which differ in the way the momenta of the colliding hadrons are used as reference directions. While in the GJ frame the polarization axis $z$ is the direction of one of the two colliding hadrons, in the CS frame such a polarization axis has the direction of the bisector of the angle formed by the direction of the incoming hadrons in the gauge boson rest frame. These two frames are related by a rotation in the production plane which is explicitly given in Appendix \ref{sec:frame transformation}.

The expressions for DMEs presented in Section \ref{sec:DMEs} apply in the GJ frame, and we start from presenting the results obtained using these formulas. In Figs. \ref{fig:DME-Z0-GJframe} and \ref{fig:DME-Wp-GJframe} we present our predictions for the six nonvanishing DMEs $\rho_i$'s, associated with the $Z^0$ and $W$ production, respectively. We show our results as a function of gauge boson transverse momentum $q_T$. 
The qualitative behaviour of all the DMEs is similar independently of the UGD used. At low $q_T$ transverse polarization dominates, while the longitudinal polarization components peaks in the ballpark of the gauge boson mass reaching about $\sim 20\%$. Regarding the interference contributions, $\rho_{\Delta \Delta}$ is similar in shape as $\rho_L$ and about a factor of two smaller as suggested by the Lam--Tung relation. Another sizeable interference contribution is $\rho_{T_p}$ which drives the lepton--antilepton forward--backward asymmetry. Notice however that it will enter with the small combination of couplings $c_Z$ for the $Z$--boson.

We have considered different models for the  proton UGD's. 
As the integrands in Eq. (\ref{eq:parton_level_xsec}) display a different $\kv$--dependence for different DMEs, one may expect a sensitivity of the magnitude and transverse momentum dependence of some of the $\rho_i$'s on the shape of $f(x,\kv)$. Such expectations are confirmed by the results presented in Figs. \ref{fig:DME-Z0-CSframe} and \ref{fig:DME-Wp-CSframe}. The more sensitive angular coefficients to the description of $f(x,\kv)$ are $\rho_{\Delta\Delta}$ and $\rho_{T_p}$. In addition, our results also indicate that the DMEs associated with the $Z^0$ and $W^+$ production are, in general, similar, differing mainly in the predictions for $\rho_\Delta$ and $\rho_{\nabla_P}$. 
We observe no sensitivity to non-linear effects, the results of KS--linear and KS--nonlinear UGDs almost overlap.

It is interesting to have a look also at the DMEs in the CS frame which is more popular in the experimental analyses. We obtain these using the transformation presented in Appendix \ref{sec:frame transformation} and show them in Figs(\ref{fig:DME-Z0-CSframe}, \ref{fig:DME-Wp-CSframe}). At small $q_T$, the DMEs should be not depend on the frame. We find that all shapes of DMEs are similar out to $q_T \sim 10 \, \rm{GeV}$. At large $q_T$ there are however large differences, for example at $q_T \sim 200 \, \rm{GeV}$ we have an equipartitioning of all three polarization states in the CS frame.
As regards the sensitivity to the UGD, none of the two frames is preferred.

\subsection{Lam--Tung relation}

\begin{figure}[t]
    \centering
    \includegraphics[width=0.45\linewidth]{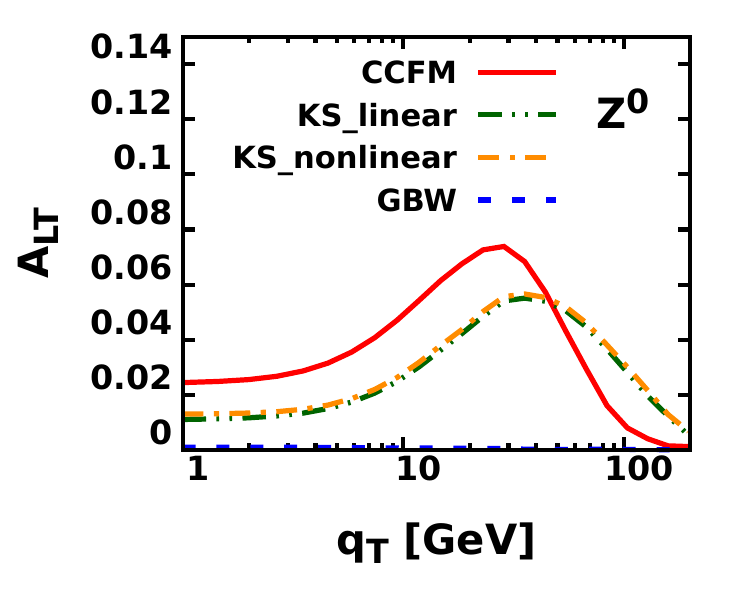}
    \includegraphics[width=0.45\linewidth]{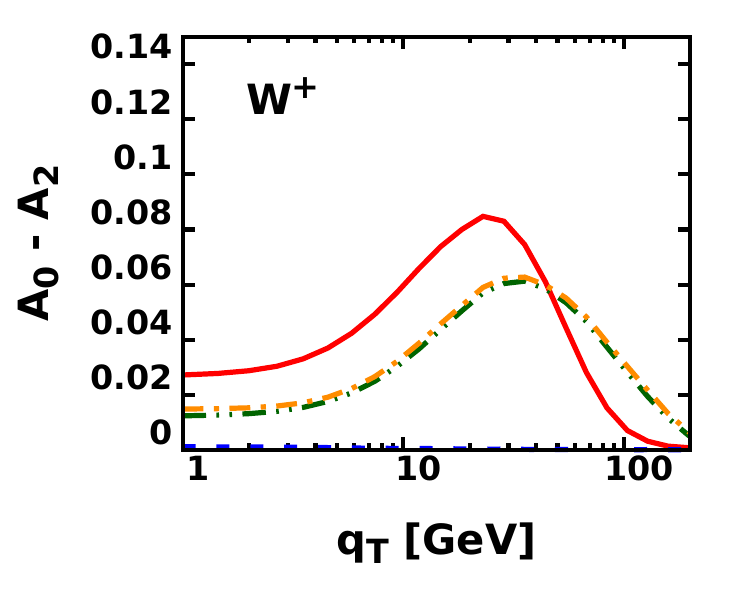}\\ 
    \caption{The Lam-Tung relation for the $Z^0$ (left panels) and $W^+$ (right panels) gauge bosons, evaluated in the forward rapidity range $2 \le y \le 4$. }
    \label{fig:LT}
\end{figure}

Finally, in Fig. \ref{fig:LT} we present our predictions for the transverse momentum dependence of the Lam--Tung relation, defined by $A_{LT} = (2W_L - 4W_{\Delta \Delta})/(2W_T + W_L)$. Notice that the combination $A_{LT}$ is invariant under rotations of the coordinate system in the production plane and hence the same in GJ and CS frames. As discussed in section \ref{sec:DMEs}, the parton model predicts $A_{LT} = 0$.  We present our predictions associated with the $Z^0$ (left panel) and $W^+$ (right panel) production 
for forward rapidities ($2.0 \le y \le 4.0$). Clearly, we predict $A_{LT} \neq 0$ in the kinematical range considered, with the results for both gauge bosons being similar. Moreover, the CCFM UGD predicts a peak at smaller values of $p_T$  in comparison with the predictions from the other UGDs. In addition, the KS-linear and KS-nonlinear predictions are almost identical, which indicates that the non-linear effects have a neglibigle impact on the Lam--Tung relation. 
For the GBW UGD the Lam--Tung relation $A_{LT}=0$ is almost fulfilled. To understand this behaviour better,
it is instructive to analyze in the massless quark limit the integration kernel relevant for the Lam--Tung relation. It reads
\begin{eqnarray}
 I_{\rm Lam-Tung}(z,\qv,z\kv) &=&   I_L(z,\qv,z\kv) - 2  I_{\Delta \Delta}(z,\qv,z\kv) \nonumber \\
 &=& |C_f^G|^2 \Big( (g^G_{V,f})^2
 + (g^G_{A,f})^2 \Big) \, \frac{4 (1-z)}{z} \, \nonumber \\
 &\times& \Big\{ (1-z)M^2 \Phi_0^2 - \bPhi^2 + 2 \frac{(\qv \cdot \bPhi)^2}{\qv^2} \Big\} \,. 
\end{eqnarray}
Expanding in the limit of  $\kv^2 \ll \qv^2 + \epsilon^2 \equiv \mu^2 $, i.e.
\begin{eqnarray}
    I_{\rm Lam-Tung}(z, \qv, z\kv) = I_0 + \frac{1}{2} z^2 \frac{\kv^2}{\mu^2} I_2 + \dots \, , 
\end{eqnarray}
one immediately finds that not only is $I_0 = 0$ [see Eq. (\ref{eq:Phi})], but also $I_2 =0$. Now, we have, 
\begin{eqnarray}
    \int d^2 \kv \, \kv^2 \, f(x,\kv) \rightarrow x g(x,\mu^2) \, , 
\end{eqnarray}
where $g(x,\mu^2)$ is the collinear gluon distribution of the target, so that in the collinear limit, the Lam--Tung structure function vanishes.
In our case a finite contribution violating the Lam--Tung relation comes from the large--$\kv$ tail of the unintegrated gluon distribution, as in the approach of \cite{Motyka:2016lta}.
We now understand why the GBW distribution, which has a gaussian behaviour in $|\kv|$ gives rise only to a small violation of the Lam--Tung relation.

\section{Summary}

\label{sec:sum}
Over the last years, the measurement and description of the Drell--Yan angular coefficients has been a subject of intense experimental and theoretical activity. Such studies were strongly motivated by the possibility to use these quantities to improve our understanding of the particle production mechanism. In this paper, we have focused on forward rapidities, where new dynamical effects can be expected to modify the description of the gauge boson production, implying the breakdown of the standard collinear factorization formalism. In particular, we have considered the color-dipole $S$-matrix framework, which was applied for the electroweak gauge boson production in our previous studies \cite{Bandeira:2024zjl,Bandeira:2024jjl}, and derived, for the first time, the corresponding  helicity DMEs for $Z$ and $W$ production. These can serve to evaluate the DY angular coefficients associated with the lepton angular distributions in $Z$ and $W$ decays. We have demonstrated that our formalism implies that the angular distribution is fully characterized by six nonvanishing angular coefficients. Moreover, we have presented results for the transverse momentum dependence of the distinct angular coefficients considering $pp$ collisions at $\sqrt{s} = 14$ TeV and that the dileptons are produced in the rapidity range $2 \le y \le 4$. Four distinct models for the proton UGD have been considered, which has allowed to estimate the sensitivity of the coefficients on the description of the QCD dynamics. Our results indicate that the predictions associated with the $Z$ and $W$ decays are similar and that the impact of the non-linear QCD effects is small.

A final comment is in order. In this paper, we have established the formalism needed to estimate the DY angular coefficients using the color - dipole $S$-matrix framework. Although we have presented predictions for the transverse momentum dependence of these coefficients, a comparison with the data was not performed since the experimental cuts assumed by the LHC collaborations were not considered. Moreover, the magnitude of the next-to-leading order corrections for our predictions is still an open question (For recent advances on this topic see, e.g., Ref. \cite{Taels:2023czt}). Both aspects deserve more detailed analysis, which we intend to perform in forthcoming studies.

\section*{Acknowledgments}
Y.B.B.  and V.P.G. were  partially supported by CNPq, CAPES (Finance code 001), FAPERGS and  INCT-FNA (Process No. 464898/2014-5). W.S. was partially supported by the Polish National Science Center Grant No. UMO-2023/49/B/ST2/03665.
\appendix

\section{Light front wave functions and helicity density matrix elements}\label{app:LFWF}

For the convenience of the reader, in this appendix we collect the explicit expression for the relevant LFWFs obtained in \cite{Bandeira:2024zjl}.

We will use that the light front wave functions in the coordinate space can be obtained by a Fourier transform of  the corresponding WF in the momentum space 
\begin{eqnarray}
\Psi_{\eta\eta'}^{(\lambda)}(z,\rr) \Bigg\vert_{V}  &=\  & \int\frac{\dd^{2}\qv}{(2\pi)^{2}}e^{-i \rr\qv} \, \Psi_{\eta\eta'}^{(\lambda)}(z,\qv) \Bigg\vert_{V} \, , \nonumber \\
&=\  & C_{f}^{G}g_{V,f}^{G}\sqrt{z(1-z)}\int\frac{\dd^{2}\qv}{(2\pi)^{2}}e^{- i\rr\qv}\frac{E^*_\mu(q,\lambda)\Gamma_V^\mu}{\qv^{2}+\epsilon^{2}},
\end{eqnarray}
and
\begin{eqnarray}
\Psi_{\eta\eta'}^{(\lambda)}(z,\rr) \Bigg\vert_{A} &=\  & \int\frac{\dd^{2}\qv}{(2\pi)^{2}}e^{-i\rr\qv}\Psi_{\eta\eta'}^{(\lambda)}(z,\qv)\Bigg\vert_{A}\, , \nonumber \\
&=\  & C_{f}^{G}g_{A,f}^{G}\sqrt{z(1-z)}\int\frac{\dd^{2}\qv}{(2\pi)^{2}}e^{-i\rr\qv}\frac{E^*_\mu(q,\lambda)\Gamma_A^\mu}{\qv^{2}+\epsilon^{2}},
\end{eqnarray}
with $\epsilon^{2}=(1-z)M^{2}+z(m_{b}^{2}-m_{a}^{2})+z^{2}m_{a}^{2}$ and
\begin{eqnarray}
    \Gamma_V^\mu &=&     \bar u(1-z,\qv,\lambda') \Big\{ \gamma^\mu  +  (m_b - m_a) \frac{k^\mu}{M^2_G} \Big \} u(1,\pmb 0, \lambda) \, \nonumber \\
    \Gamma_A^\mu &=&    \bar u(1-z,\qv,\lambda') \Big\{ \Big(\gamma^\mu  + (m_a + m_b)\frac{k^\mu}{M^2_G} \Big) \gamma_5   \Big \} u(1,\pmb 0, \lambda) \, ,
\end{eqnarray}
where $m_a$ and $m_b$ are the masses of the quarks before and after the gauge boson emission, and the polarization vectors to be used are  
\begin{eqnarray}
    E_\mu(q,\pm 1) = E_\mu^{\perp}(\pm 1) - \frac{E^{\perp}(\pm 1)\cdot q}{q^+} n^-_\mu \, ,\quad   E_\mu(q,0) = - \frac{M_G}{q^+} \, n^-_\mu \, . 
\end{eqnarray}


Then the LFWF, understood as a $2 \times 2$ matrix in the space of (anti--)quark polarizations, reads
\begin{eqnarray}
\Psi^{(\lambda)}(z,\kv) = C_f^G  \Big\{ g^G_{V,f} \Psi_V^{(\lambda)}(z,\kv) + g^G_{A,f} \Psi_A^{(\lambda)}(z,\kv) \Big\} \, , 
\end{eqnarray}
with 
\begin{align}
\begin{split}
    \Psi_{V}^{(\pm)}(z,\boldsymbol{k})=\  & \frac{C_{f}^{G}g_{V,f}^{G}\sqrt{z}}{\kv^2 + \epsilon^2} \chi_{\eta'}^{\dagger}\Big\{ \frac{2-z}{z}\left(\boldsymbol{k}\cdot\be^{*}(\pm)\right)\openone 
    + i\left(\boldsymbol{k}\times\be^{*}(\pm)\right)_{\hat{z}}\sigma_{3}
    \\ & 
    -\Gamma_{V}\left(\boldsymbol{\sigma}\cdot\be^{*}(\pm)\right)\sigma_{3} \Big\} \chi_{\eta}\label{eq:MS-LFWF-V+-}    
\end{split}\\
\Psi_{V}^{(0)}(z,\boldsymbol{k})=\  & \frac{C_{f}^{G}g_{V,f}^{G}}{\sqrt{z}M [\kv^2 + \epsilon^2]} \chi_{\eta'}^{\dagger}\left\{ \Lambda_{V}\openone+z(m_{b}-m_{a})\left(\boldsymbol{\sigma}\cdot\boldsymbol{k}\right)\sigma_{3}\right\} \chi_{\eta}\label{eq:MS-LFWF-V0}\\
\begin{split}
\Psi_{A}^{(\pm)}(z,\boldsymbol{k})=\  & \frac{C_{f}^{G}g_{A,f}^{G}\sqrt{z}}{\kv^2 + \epsilon^2} \chi_{\eta'}^{\dagger}\Big\{ \frac{2-z}{z}\left(\boldsymbol{k}\cdot\be^{*}(\pm)\right)\sigma_{3}
+i\left(\boldsymbol{k}\times\be^{*}(\pm)\right)_{\hat{z}}\openone
\\ &
-\Gamma_{A}\left(\boldsymbol{\sigma}\cdot\be^{*}(\pm)\right)\Big\} \chi_{\eta}\label{eq:MS-LFWF-A+-}    
\end{split}
\\
\Psi_{A}^{(0)}(z,\boldsymbol{k})=\  & \frac{C_{f}^{G}g_{A,f}^{G}}{\sqrt{z}M [\kv^2 + \epsilon^2]} \, \chi_{\eta'}^{\dagger}\left\{ -\Lambda_{A}\sigma_{3}-z(m_{b}+m_{a})\left(\boldsymbol{\sigma}\cdot\boldsymbol{k}\right)\openone \right\} \chi_{\eta} .
\label{eq:MS-LFWF-A0}
\end{align}


\section{Frame transformation}
\label{sec:frame transformation}
In Eq.~\eqref{eq:angular_distribution} we present the full expression for the angular distribution in terms of density matrix elements, however using the Eq.~\eqref{eq:structure_func_def} one can connect the density matrix elements with the structure functions elements. Therefore, the angular distribution in terms of the structure functions at a given production plane frame $F$ is:
\begin{eqnarray}\label{appx:lep-ang-dist}
    \frac{d N}{d \Omega} &=&  \frac{3}{8\pi} \frac{1}{2W_T + W_L}
    \Bigg[ g_T W_T + g_L W_L + g_\Delta W_\Delta  + g_{\Delta\Delta} W_{\Delta\Delta} \nonumber \\ 
    & & + c_G g_{T_P} W_{T_P} + c_G g_{\nabla_P} W_{\nabla_P} + c_G g_\nabla W_\nabla + g_{\Delta\Delta_P} W_{\Delta\Delta_P} + g_{\Delta_P} W_{\Delta_P}
    \Bigg]\, ,
\end{eqnarray}
where 
\begin{eqnarray}
    g_T = 1 + \cos^2\theta \, , \quad & g_L = 1 - \cos^2\theta \, , \quad & g_{T_P} = 2 \cos\theta \, , \nonumber\\ 
    g_{\Delta\Delta} = \sin^2\theta\cos 2\phi \, , \quad & g_\Delta = \sin 2\theta\cos\phi \, , \quad & g_{\nabla_P} = 2\sin\theta\cos\phi \, , \nonumber\\
    g_{\Delta\Delta_P} = \sin^2\theta\sin 2\phi \, , \quad & g_{\Delta_P} = \sin 2\theta\sin\phi \, , \quad & g_{\nabla} = 2 \sin\theta\sin\phi \, . 
\end{eqnarray}
\par 
We can parameterize the transformation from one observation frame to another by a single angle, describing a rotation around the $y$ axis. The rotation matrix 
\begin{eqnarray}
    R_y(\gamma) &=& 
    \begin{pmatrix}
        \cos\gamma &  0 & - \sin \gamma \\ 
        0 & 1 & 0 \\ 
        \sin\gamma & 0 & \cos\gamma
    \end{pmatrix}
\end{eqnarray}
transforms the components of a vector, reproducing the effect of a rotation of the $Z$ and $x$ axis in the production plane. The coordinates of the unit vector indicating the positive lepton direction in the \emph{old} frame, 
\begin{eqnarray}
    \hat r &=& 
    \left( 
        \sin\theta\cos\phi,\,\sin\theta\sin\phi,\,\cos\theta 
    \right)\,,
\end{eqnarray}
can be expressed as a function of the coordinates in the new frame as 
\begin{eqnarray}\label{appx:coor-rel}
    \sin\theta\cos\phi &=& \cos\gamma\sin\theta'\cos\phi'+\sin\gamma\cos\theta' \, , \\
    \sin\theta\sin\phi &=& \sin\theta\sin\phi  \, , \\ 
    \cos\theta &=& -\sin\gamma\sin\theta'\cos\phi' + \cos\gamma\cos\theta' \, .
\end{eqnarray}

Substituting Eq.~\eqref{appx:coor-rel} into Eq.~\eqref{appx:lep-ang-dist}, we obtain the angular distribution in the rotated frame
\begin{eqnarray}\label{appx:rot-lep-ang-dist}
    \frac{d N}{d \Omega} &=&  \frac{3}{8\pi} \frac{1}{2W_T + W_L} \Bigg\{ 
    g_T
        \Big[
            \Big(1- {\sin^2\gamma \over 2}\Big) W_T + {\sin^2\gamma \over 2} W_L + {\sin2\gamma \over 2} W_\Delta + {\sin^2 \gamma \over 2}W_{\Delta\Delta}
        \Big] \nonumber \\ 
&+&
    g_L 
        \Big[
            \sin^2\gamma W_T + \cos^2\gamma W_L - \sin2\gamma W_\Delta - \sin^2\gamma W_{\Delta\Delta}
        \Big] \nonumber \\ 
&+&
    g_{\Delta} 
        \Big[ 
            -{\sin2\gamma \over 2}W_T + {\sin2\gamma \over 2}W_L + \cos2\gamma W_\Delta + {\sin2\gamma \over 2}W_{\Delta\Delta}
        \Big] \nonumber \\ 
&+&
    g_{\Delta\Delta}
        \Big[ 
            {\sin^2\gamma \over 2}W_T - {\sin^2\gamma \over 2}W_L - {\sin2\gamma \over 2}W_{\Delta} + \Big({1 - \sin^2\gamma \over 2}\Big)W_{\Delta\Delta}
        \Big] \nonumber \\ 
&+&
    c_G g_{T_p}
        \Big[ 
            \cos\gamma W_{T_p} + \sin\gamma W_{\nabla_p}
        \Big] \nonumber \\
&+& 
    c_G g_{\nabla_p} 
        \Big[ 
            -\sin\gamma W_{T_p} + \cos\gamma W_{\nabla_p}
        \Big]\nonumber \\ 
&+&
   g_{\Delta\Delta_p} 
        \Big[ 
            \cos\gamma W_{\Delta\Delta_p} - \sin\gamma W_{\Delta_p}
        \Big]\nonumber \\ 
&+&
   g_{\Delta_p} 
        \Big[ 
            \sin\gamma W_{\Delta\Delta_p} + \cos\gamma W_{\Delta_p}
        \Big]\nonumber \\         
&+&
    c_G g_{\nabla} W_{\nabla} \Bigg\}\,.
\end{eqnarray}

Therefore, comparing Eq.~\eqref{appx:lep-ang-dist} and Eq.~\eqref{appx:rot-lep-ang-dist}, we obtain the relation between the structure function in different frames.

To connect the GJ frame with the CS frame, the angular relation is 
\begin{eqnarray}
    \cos\gamma = { 1 \over  \sqrt{1 + \beta^2}} \, , \hskip3em \sin\gamma = {\beta \over \sqrt{1 + \beta^2 }} \, 
\end{eqnarray}
for $\beta = |\qv| / M$. Therefore, 

\begin{eqnarray}
\resizebox{1.0\hsize}{!}{$
    \begin{pmatrix}
        W_T \\ W_L \\ W_\Delta \\ W_{\Delta\Delta} \\ W_{T_p} \\ W_{\nabla_p} \\ W_{\Delta\Delta_p} \\ W_{\Delta_p} \\ W_\nabla
    \end{pmatrix}_{CS}
    = 
    { 1 \over 1 + \beta^2}
    \begin{pmatrix}
    1 +  {\beta^2 \over 2} &  {\beta^2 \over 2} &  \beta &  {\beta^2  \over 2} 
    & 0 & 0 & 0 & 0 & 0 \\   
    \beta^2 & 1 & - 2\beta & - \beta^2
    & 0 & 0 & 0 & 0 & 0 \\   
    -\beta & \beta & 1 - \beta^2 & \beta
     & 0 & 0 & 0 & 0 & 0 \\   
     {\beta^2 \over 2} & - {\beta^2 \over 2} & - \beta & 1 + {\beta^2 \over 2}
     & 0 & 0 & 0 & 0 & 0 \\   
     0 & 0 & 0 & 0 & 
     \sqrt{1+\beta^2} & \beta\sqrt{1+\beta^2}   
     & 0 & 0 & 0 \\
     0 & 0 & 0 & 0 & 
     -\beta\sqrt{1+\beta^2} & \sqrt{1+\beta^2}
     & 0 & 0 & 0 \\
     0 & 0 & 0 & 0 & 0 & 0 & 
     \sqrt{1+\beta^2} & -\beta\sqrt{1+\beta^2}
      & 0 \\
      0 & 0 & 0 & 0 & 0 & 0 & 
     \beta\sqrt{1+\beta^2} &  \sqrt{1+\beta^2}
      & 0 \\
      0 & 0 & 0 & 0 & 0 & 0 & 0 & 0 & 1+\beta^2
    \end{pmatrix}
    \begin{pmatrix}
        W_T \\ W_L \\ W_\Delta \\ W_{\Delta\Delta} \\ W_{T_p} \\ W_{\nabla_p} \\ W_{\Delta\Delta_p} \\ W_{\Delta_p} \\ W_\nabla
    \end{pmatrix}_{GJ} $}\, . 
\end{eqnarray}

\section{The angular coefficients}\label{appx-sec:angular-coeff}
In Eq.~\eqref{eq:ang-dist-A} we present the angular distribution parametrization, nevertheless one need connects it with the angular distribution which given us the $A_i$ ($i=0,1,2,3,4,5,6,7$) values in terms of the structure functions $W_j$ ($j=T,L,\Delta,\Delta\Delta,T_P,\nabla_P,\nabla,\Delta\Delta_P,\Delta_P$). In what follows, we'll show how to obtain such connection. Therefore, starting from the angular distribution which is written as
\begin{eqnarray}
    \frac{d N}{d \Omega} &=&  \frac{3}{8\pi} \frac{1}{2W_T + W_L}
    \Bigg[ g_T W_T + g_L W_L + g_\Delta W_\Delta  + g_{\Delta\Delta} W_{\Delta\Delta} \nonumber \\ 
    & & + c_G g_{T_P} W_{T_P} + c_G g_{\nabla_P} W_{\nabla_P} + c_G g_\nabla W_\nabla + g_{\Delta\Delta_P} W_{\Delta\Delta_P} + g_{\Delta_P} W_{\Delta_P}
    \Bigg]\, ,
\end{eqnarray}
where 
\begin{eqnarray}
    g_T = 1 + \cos^2\theta \, , \quad & g_L = 1 - \cos^2\theta \, , \quad & g_{T_P} = 2 \cos\theta \, , \nonumber\\ 
    g_{\Delta\Delta} = \sin^2\theta\cos 2\phi \, , \quad & g_\Delta = \sin 2\theta\cos\phi \, , \quad & g_{\nabla_P} = 2\sin\theta\cos\phi \, , \nonumber\\
    g_{\Delta\Delta_P} = \sin^2\theta\sin 2\phi \, , \quad & g_{\Delta_P} = \sin 2\theta\sin\phi \, , \quad & g_{\nabla} = 2 \sin\theta\sin\phi \, . 
\end{eqnarray}
\par 
Multiplying the angular distribution by $(W_T +W_L)/(W_T +W_L)$, \ one gets
\begin{eqnarray}\label{eq:theo-parametrization}
    \frac{d N}{d \Omega} &=&  \frac{3}{4\pi} \frac{1}{\lambda+3}
    \Bigg[
        1 + \lambda\cos^2\theta + \mu\sin2\theta\cos\phi + {\nu \over 2}\sin^2\theta\cos2\phi 
\nonumber \\ &&
        + \tau\sin\theta\cos\phi + \eta\cos\theta + \xi\sin^2\theta\sin2\phi
        + \zeta\sin2\theta\sin\phi + \chi\sin\theta\sin\phi
    \Bigg]\, ,
\end{eqnarray}
where 
\begin{eqnarray}
    \lambda = {W_T - W_L \over W_T + W_L} \, , \, \,  \mu = {W_\Delta \over W_T + W_L}\, , \, \, \nu = {2W_{\Delta\Delta} \over W_T + W_L} \, , \, \, 
    \tau = {2c_{G} W_{\nabla_P} \over W_T + W_L}\, , 
\nonumber \\
    \eta = {2c_GW_{T_P} \over W_T + W_L} \, ,  \, \, \xi = {W_{\Delta\Delta_P} \over W_T + W_L}\, , \, \,
    \zeta = {W_{\Delta_P} \over W_T + W_L} \, , \, \, \chi = {2c_GW_\nabla \over W_T + W_L}\, .   
\end{eqnarray}

The Eq.~\eqref{eq:theo-parametrization} is an usual parametrization for the angular distribution, quite often present in theoretical papers (see for instance \cite{Boer:2006eq,Lyubovitskij:2024jlb}). However, the most used parametrization found in experimental works is the one presented in Eq.~\eqref{eq:ang-dist-A}. Therefore, by comparison, one has that these two parametrization connects trough
\begin{eqnarray}
     A_0 = { 2(1-\lambda) \over \lambda + 3 } = {2W_L \over 2W_T + W_L}\, , \,\, &&\, \,
     A_1 =  {4\mu \over \lambda + 3} = {2W_\Delta \over 2W_T + W_L} \, , \,\,
 \nonumber \\ 
     A_2 = {4\nu \over \lambda + 3} = {4W_{\Delta\Delta} \over 2W_T + W_L}\, , \, \, &&\, \,
     A_3 = {4\tau \over \lambda + 3} = {4 c_G W_{\nabla_P} \over 2W_T + W_L}\, , \, \, 
 \nonumber \\ 
     A_4 = {4\eta \over \lambda + 3} = {4 c_G W_{T_P} \over 2W_T + W_L}\, , \, \, &&\, \,
     A_5 = {4\xi \over \lambda + 3} = {2 W_{\Delta\Delta_P} \over 2W_T + W_L} \, , \, \, 
 \nonumber \\ 
    A_6 = {4\zeta \over \lambda + 3} = {2W_{\Delta_P} \over 2W_T + W_L}\, , \, \, &&\, \,
    A_7 = {4\chi \over \lambda + 3} = {4 c_G W_\nabla \over 2W_T + W_L}\, .
\end{eqnarray}
where we have used that 
\begin{eqnarray}
    {4 \over \lambda + 3}\left( 1 + \lambda \cos^2\theta \right) &=&
    {1 \over \lambda + 3}\left( 4+ (\lambda + 3\lambda)\cos^2\theta \right)\, ,
\nonumber \\ 
    {4 \over \lambda + 3}\left( 1 + \lambda \cos^2\theta \right) &=&
    {1 \over \lambda + 3}\left( 4 +\lambda - \lambda  + (\lambda + 3\lambda + 3 - 3)\cos^2\theta \right)\, ,
\nonumber \\ 
    \therefore {4 \over \lambda + 3}\left( 1 + \lambda \cos^2\theta \right) &=&
    1 + \cos^2\theta + {(1-\lambda)\over \lambda + 3}(1-3\cos^2\theta)\, .
\end{eqnarray}

\bibliography{references.bib}

\end{document}